\documentclass[%
 reprint,
superscriptaddress,
 amsmath,amssymb,
prb,
]{revtex4-1}

\usepackage{graphicx}
\usepackage{dcolumn}
\usepackage{bm}
\usepackage{comment}

\usepackage{braket}
\usepackage{color}
\usepackage{array}
\newcolumntype{?}{!{\vrule width 1pt}}

\begin{document}


\title{{Theory of magneto-optical properties of neutral and charged excitons in GaAs/AlGaAs quantum dots}}





 
\author{Diana Csontosov\'{a}}%
    \affiliation{Department of Condensed Matter Physics, Faculty of Science, Masaryk University, Kotl\'a\v{r}sk\'a~267/2, 61137~Brno, Czech~Republic}
    \affiliation{Czech Metrology Institute, Okru\v{z}n\'i 31, 63800~Brno, Czech~Republic}
  

    
    
\author{Petr Klenovsk\'{y}}%
    \email{klenovsky@physics.muni.cz}
    \affiliation{Department of Condensed Matter Physics, Faculty of Science, Masaryk University, Kotl\'a\v{r}sk\'a~267/2, 61137~Brno, Czech~Republic}
    \affiliation{Czech Metrology Institute, Okru\v{z}n\'i 31, 63800~Brno, Czech~Republic}

\begin{abstract}

Detailed theoretical study of the magneto-optical properties of weakly confining GaAs/AlGaAs quantum dots is provided. We focus on the diamagnetic coefficient and the $g$-factor of the neutral and the charged excitonic states, respectively, and their evolution with various dot sizes for the magnetic fields applied along $[001]$ direction. For the calculations we utilize the combination of ${\bf k}\cdot{\bf p}$ and the configuration interaction methods. We decompose the theory into four levels of precision,~i.e., (i) single-particle electron and hole states, (ii) non-interacting electron-hole pair, (iii) electron-hole pair constructed from the ground state of both quasiparticles and interacting via the Coulomb interaction (i.e. with minimal amount of correlation), and (iv) that including the effect of correlation. The aforementioned approach allows us to pinpoint the dominant influence of various single- and multi-particle effects on the studied magneto-optical properties, allowing the characterization of experiments using models which are as simple as possible, yet retaining the detailed physical picture.

 \end{abstract}

\pacs{Valid PACS appear here}

\maketitle

\section{Introduction}

Semiconductor III-V quantum dots (QDs) have been extensively studied in the past, owing to their properties stemming from the zero-dimensional nature of the quantum confinement. Those are,~e.g.,~an almost $\delta$-function-like emission spectra, which lead to a number of appealing applications in semiconductor opto-electronics. Hence, such QDs are crucial for classical telecommunication devices as low threshold/high bandwidth semiconductor lasers and amplifiers~\citep{Bimberg1997,Ledentsov,Heinrichsdorff1997,Schmeckebier2017, Unrau_laserphotonics_2014}, as sources of single and entangled photon pairs that might be used for the quantum communication~\cite{yuan_electrically_2002, martin-sanchez_single_2009, salter_entangled-light-emitting_2010,takemoto_quantum_2015, schlehahn_single-photon_2015, kim_two-photon_2016, paul_single-photon_2017, muller_quantum_2018, Plumhof2012,Trotta:16, Aberl2017, Klenovsky2018}, or other quantum information technologies \cite{li_all-optical_2003, robledo_conditional_2008, kim_quantum_2013, yamamoto_present_2011, michler_quantum_2017, Krapek2010, Klenovsky2016, Kindel_prb2010, Sala2018, SteindlPL2019}. 
However, the aforementioned applications are mostly based on the In(Ga)As QDs embedded in GaAs matrix. In that material system, the QDs are compressively strained due to the lattice mismatch between InAs and GaAs of $\sim 7\,\%$~\cite{landoltbornstein}. That in conjunction with the lack of inversion symmetry in the zincblende semiconductors, leads to considerable shear strain in and around the dots, causing among others the non-negligible fine-structure-splitting (FSS) of the QD ground-state exciton doublet~\cite{Trotta:16,Aberl2017,Klenovsky2018}. As a result, the emitted photons are distinguishable, hampering their use,~e.g., as sources of single entangled states for quantum key distribution (QKD) protocols~\cite{Bennett1993,Ekert1991}.

To overcome that drawback, recently GaAs QDs embedded in $\rm Al_{0.4}Ga_{0.6}As$ matrix were fabricated by the droplet-etching method~\cite{Wang2007,Heyn2009,Huo:APL2013}. Since the lattice mismatch in that material system is only $\sim 0.06\%$, they show very small FSS as was recently confirmed in Refs.~\cite{Schimpf:s,Huo2014}. Hence, because of their favorable properties, the GaAs/$\rm Al_{0.4}Ga_{0.6}As$ QDs emerged as a promising source of non-classical states of light, such as single photons with a strongly suppressed multi-photon emission probability~\cite{doi:10.1063/1.5020038}, highly indistinguishable photon states~\cite{Huber2017,Reindl2019,doi:10.1021/acs.nanolett.8b05132,Liu2019}, and single polarization entangled photon-pairs with an almost near unity degree of entanglement~\cite{Huber2017,Keil2017,Huber2018,Gurioli2019}.

Because of their well defined shape and size, the almost negligible built-in strain and alloy disorder, GaAs/$\rm Al_{0.4}Ga_{0.6}As$ QDs are excellent system for testing the current quantum mechanical theory of nanostructures. A preferred way of doing so is the comparison between the experimentally measured and theoretically predicted properties (emission energy, oscillator strength, polarization) for QDs under externally applied perturbations. Those might be strain, electric, or magnetic fields and we have recently shown~\cite{Huber2019} the inadequacy of the single-particle model~\cite{Bayer2002,Witek2011} for the description of the latter. Further studies~\cite{Lobl2019} recently demonstrated the QD size dependence of the applied magnetic field response for excitons and charged trions. However, the detailed theoretical description of that is still missing and we fill that gap in this paper.

The paper is organized as follows: we start with the description of the theory model in section II, continue with discussion of the theory of the response of GaAs/AlGaAs QDs to externally applied magnetic field with particular emphasis on the diamagnetic shift and $g$-factor, and in section III we further compare our results with available experiments. Thereafter in section IV we focus on the magnetic field response of charged trions and we conclude in section V.

\section{Theory model and studied quantum dot}

\begin{figure*}[htbp]
	\centering
    \includegraphics [width=140mm]{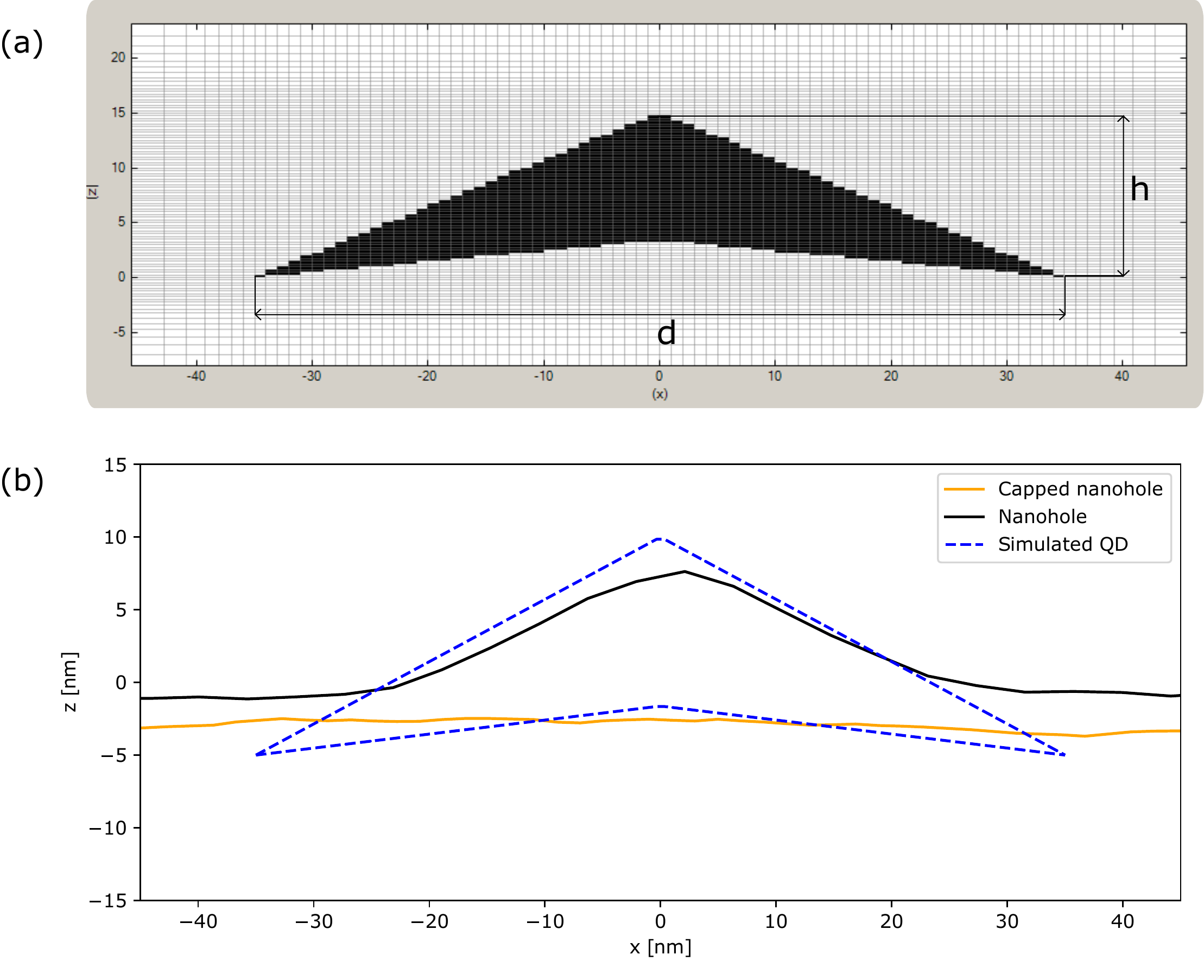}
      \caption{(a) Side view of computed $\rm GaAs$ QD (black object) embedded in $\rm Al_{0.4}Ga_{0.6}As$ matrix. Horizontal and vertical lines mark the grid of the simulation space and the scale for $(x)$ and $(z)$ direction is given in units of $\rm nm$; $h$ and $d$ label the height and the base diameter of QD, respectively. The shape of QD was inspired by Ref.~\cite{Huber2019}, where it was chosen to fit the atomic force microscopy (AFM) measurements and optimized to match the emission energy and magneto-optical properties of X$^0$. (b) Comparison of our simulated dot with AFM line scans of an exemplary $\rm GaAs/Al_{0.4}Ga_{0.6}As$ QD. The AFM measurement of the nanohole (black solid curve) was performed by the scientists at the Johannes Kepler University in Linz and the measurement of filled nanoholes (orange solid curve) was performed by scientists at the Czech Metrology Institute in Brno. Broken curve shows the simulated QD. Note, that in order to facilitate comparison with the computed structure we turned the measured AFM linescans upside-down in (b). Note, that since the actual thickness of the GaAs capping layer cannot be resolved by AFM measurements, we set the leftmost value of black and orange curves in the left part of the graph in panel (b) to zero (not shown).}
    \label{fig1}
\end{figure*} 

We theoretically study the excitonic structure of the GaAs/$\rm Al_{0.4}Ga_{0.6}As$ QDs using the following methodology. It starts with the implementation of the 3D QD model structure (size, shape, chemical composition), see Fig.~\ref{fig1}, and carries on with the calculation of the strain and the piezoelectricity. The resulting strain and polarization fields then enter the eight-band $\mathbf{k}\cdot\mathbf{p}$ Hamiltonian~\cite{Klenovsky2019}. Thereafter, for the QD with applied magnetic field, the Hamiltonian introduced in Ref.\cite{Luttinger1956} with added Pauli term describing the interaction between the magnetic field and the spin is solved~\cite{t_Andlauer} using the Nextnano suite~\cite{Birner:07} yielding the electron and hole single-particle (SP) states. For the full list of material parameters used in this work see Ref.~\cite{SupMatParam}~(see, also, references~\cite{Luttinger1956,Varshni1967,Dresselhaus1955,Vurgaftman2001,deGironcoli1989,Wei1998,Klenovsky2013} therein). The Coulomb interaction between the quasiparticles and the correlation is accounted for by employing the configuration interaction (CI) method~\cite{Klenovsky2017,Klenovsky2019}. Using the theory toolbox described thus far, we obtain the eigenenergies and eigenfunctions of various complexes like the neutral (X$^0$), the positively (X$^+$), and the negatively (X$^-$) charged exciton, see Fig.~\ref{figProbab} for the probability densities of those states. See also Appendix I and Appendix II for details about the CI computation method and the evaluation of the corresponding results. Note, that since our theoretical description is supposed to be applicable for explanation of experiments, we simulate the structure for finite temperature of $T = 8\,\rm K$.

Finally, we note that we choose the 8-band $\mathbf{k}\cdot\mathbf{p}$ model (i) on the grounds of its simplicity and (ii) since it was widely used so far to study the GaAs/AlGaAs system, see, Refs.~\cite{Wang2009,Huo2014}. Furthermore, the strain fields and piezoelectricity, considered in our model, in turn causes the SP wavefunctions to have the correct C2v symmetry, and not C4v inherent to $\mathbf{k}\cdot\mathbf{p}$ approximation. Moreover, the 8-band $\mathbf{k}\cdot\mathbf{p}$ SP solver nextnano was in the past tuned by its creators~\cite{Birner:07} to produce similar results as more elaborate methods like pseudopotentials as those of,~e.g., calculations of Bester~et~al., see Ref.~\cite{Bester2008} or other $\mathbf{k}\cdot\mathbf{p}$ solvers like those of Stier~et~al., see Ref.~\cite{Stier1999}. Furthermore, while it is known that the error of eigenenergies of electrons and holes computed by $\mathbf{k}\cdot\mathbf{p}$ method might be of the order of meV, since FSS results from CI calculations, the uncertainty of that is on the order of sub-$\mu$eV level. Note, that further slight uncertainties related to SP basis states computed by $\mathbf{k}\cdot\mathbf{p}$ are in our experience somewhat corrected by including larger basis set of CI.

The magnetic flux density ($B$) induces circulating current which leads to the magnetic momentum ($\mu$) opposed to $B$. The interaction between $B$ and $\mu$ causes, among others, the energy shift ($\Delta E$) of the spin degenerate state~\cite{vanBree2012}. In the first approximation one obtains
\begin{equation}
\Delta E = \gamma B^2,
\label{eq2.1}
\end{equation}
where $\gamma$ is the diamagnetic coefficient which is proportional to the spatial expansion of the wave function in the direction perpendicular to $B$. Hence, $\gamma$ for a carrier in a semiconductor satisfies~\cite{vanBree2012}
\begin{equation}
\gamma \propto \frac{\left\langle r^2\right\rangle}{m^*},
\label{eq2.2}
\end{equation}
where $\left\langle r^2\right\rangle$ is the average expansion of the wave function in the direction perpendicular to $B$ and $m^*$ is the effective mass of the charged carrier. Thus,~e.g., by inspection of Fig.~\ref{figProbab} one can anticipate larger $\gamma$ for X$^+$ and X$^-$ states compared to those of X$^0$.

The other dominant process which is observed is the Zeeman effect, which is due to the interaction of $B$ with the projection of the spin momentum ($S$) to the direction parallel to $B$. In the case of $B$ applied in the direction of QD growth ($B_z$), spin degeneracy of states is lifted. However, $B$ applied in the plane of QD ($B_x$) also breaks the symmetry of the system and, thus, the coupling between different states is involved in that case, like,~e.g., that between the dark with total angular momentum $\ket{\pm 2}$ and the bright with total angular momentum $\ket{\pm 1}$ states of X$^0$, respectively~\cite{vanBree2012,Bayer2002,Huber2019}. The splitting of the energy levels depends on $B$ linearly in the first approximation, and the slope of that is commonly called the $g$-factor. We note, that in the following text we focus only on $B$ applied in the growth direction,~i.e., we study the response to $B_z$. 

The aforementioned effects are observable for a variety of quasiparticles like the holes, the excitons, or other complexes. To extract $\gamma$ and $g$-factor of computed (multi-)particle complexes we use the following model which is suitable also for evaluation of the experiments~\cite{Huber2019},
\begin{equation}
E_{\uparrow / \downarrow} = E_0 + \gamma B^2 \pm \frac{1}{2} \sqrt{E_{\rm FSS}^2 + g_0^2 \mu_B^2 B^2},
\label{eq2.3}
\end{equation}
where $\uparrow / \downarrow$ labels the spin of the energy levels, $E_0$ and $E_{\rm FSS}$ are the emission and FSS energies, respectively, of the corresponding state for $B=0\,\rm{T}$; $g_0$ denotes the $g$-factor, and $\mu_B$ is the Bohr magneton. Note, that since the splitting is strongly linear in our calculations, we take into account only the zeroth term $g_0$ of $g$-factor and neglect the second order perturbation term $g_2$ introduced in Ref.~\cite{vanBree2012}.

Note, that the values of magneto-optical properties extracted from Eq.~\eqref{eq2.3} are in further text calculated for magnetic fields in the range $B \in [-0.4, 0.4]\rm \, T$.

Finally, we stress that the ``spin" is generally not a good quantum number that can be used to classify the energy states of our quasiparticles in the following. That is since (i) sizeable spin-orbit coupling is present in our system on bulk level and (ii) the quantum states calculated by CI and even by the envelope method based on multi-band $\mathbf{k}\cdot \mathbf{p}$ approximation, are composed of the single-spin states, which are mixed with different contents. This leaves us to classify the states only using the time-reversal symmetry as Kramers doublets.

\section{Size dependence of magneto-optical properties of neutral exciton}\label{sec2.3}

\begin{figure*}[htbp]
	\centering
		\includegraphics[width=160mm]{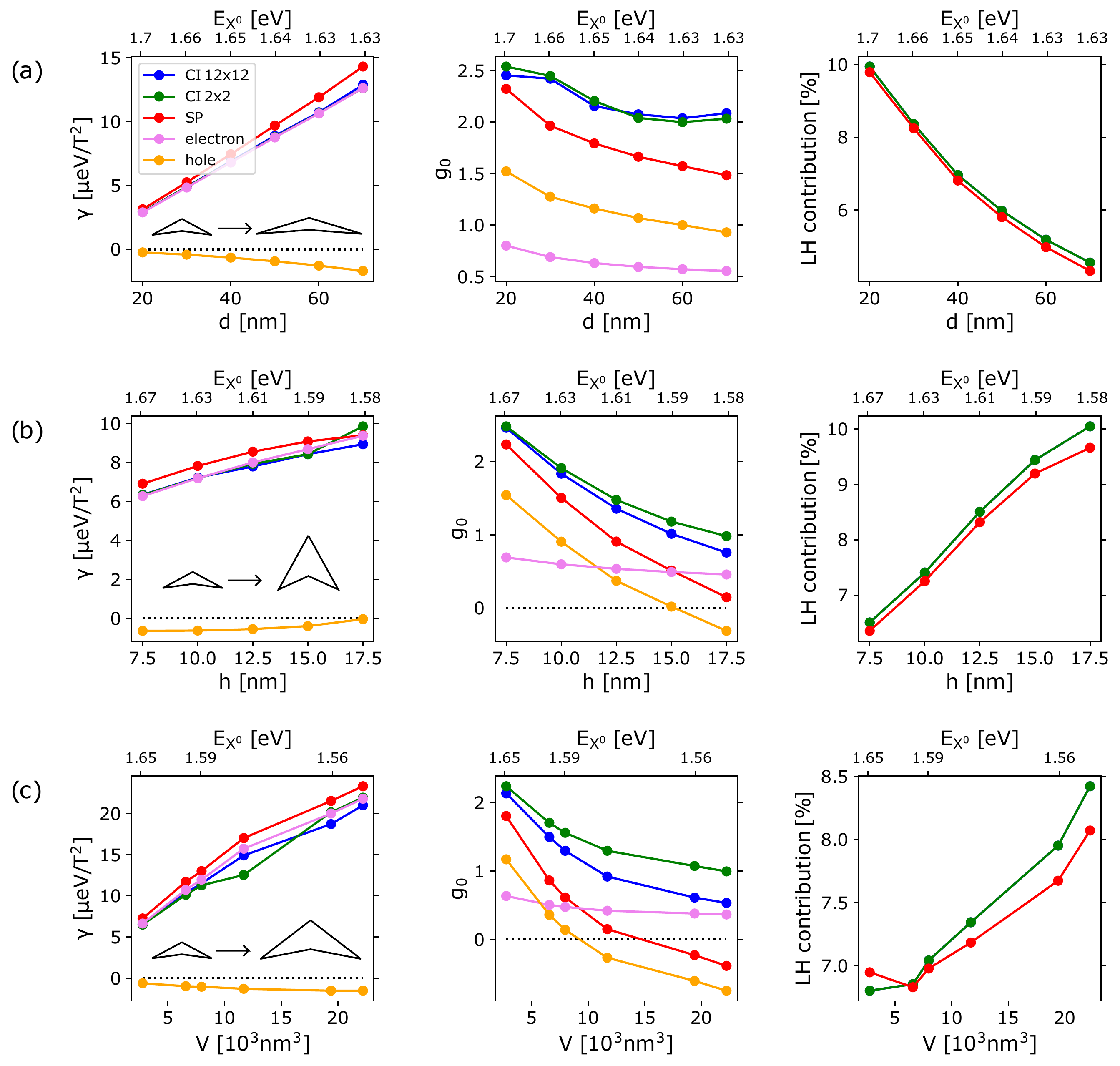}
	\caption{Dependencies of $\gamma$, $g$-factor, and LH contribution in X$^0$ ground state, respectively, for $B_z = 0\,\rm T$ on the emission energy for (a) fixed height of $h = 9\,\rm nm$, (b) fixed base diameter of $d = 40\,\rm nm$, and (c) fixed aspect ratio of $h/d = 3/13$, where $V$ is the volume of QD. The upper horizontal axis shows the emission energy of X$^0$ calculated in $12 \times 12$ basis. The insets in the first column of (a), (b), and (c) sketch the corresponding change of QD size and shape. Note, that the blue and green circles, and curves overlap in the rightmost column of graphs.} 
	\label{fig2}
\end{figure*}

In this section we study the size dependencies of the magneto-optical properties of X$^0$ ground state of the QDs shown in Fig.~\ref{fig1}. We mark the height and the diameter of QD base $h$ and $d$, respectively, see Fig.~\ref{fig1}~(a), and we track $\gamma$, $g_0$, and the light-hole (LH) content for (i) QD with fixed $h$ and varying $d$, (ii) fixed $d$ and varying $h$, and (iii) for fixed $h/d$ ratio, thus, we track the variation of QD volume ($V$) in the latter case. The dependencies of $\gamma$ and $g$-factor are shown in Fig.~\ref{fig2}. We plot the results of CI calculation for the CI basis of $12$ SP electron and $12$ SP hole states marked as $12 \times 12$ (blue) and for the basis of $2$ SP electron and $2$ SP hole states marked as $2 \times 2$ (green). The comparison between those bases allows us to study the effects of correlation. To see the effect of the Coulomb and the exchange interaction, the dependencies of SP electron--hole pair (red), electron (pink) and hole (orange) states are included in Fig.~\ref{fig2} as well.

The emission energy ($E_{\rm{X}^0}$) of X$^0$ calculated in the $12 \times 12$ basis changes for all kinds of the studied size variations in the range of $\approx 100\,\rm meV$. To be able to quantitatively compare the computed dependencies, we define the parameter $b$ as
\begin{equation}
    b = \left| \frac{f(E_{\rm{X}^0}^f) - f(E_{\rm{X}^0}^i)}{E_{\rm{X}^0}^f - E_{\rm{X}^0}^i}\right|,
    \label{eq:ParameterB}
\end{equation}
where $f(E_{\rm{X}^0}^f)$ $\{f(E_{\rm{X}^0}^i)\}$ is the studied quantity (e.g.~magneto-optical properties) for the final \{initial\} value of the size dependence of $E_{\rm{X}^0}$. Note, that since the studied dependencies are not always linear and $E_{\rm{X}^0}$ is not the quantity describing SP hole and electron states, $b$ provides only an estimation of the slope of the corresponding dependency.

\subsection{Diamagnetic coefficients} \label{sec2.3.1}

First, we study the size dependence of $\gamma$. From Eq.~\eqref{eq2.2} we expect the sensitivity of $\gamma$ to variation of $d$. This is confirmed by the numerical calculation in Fig.~\ref{fig2}~(a). We see that the absolute value of $\gamma$ for electrons ($\gamma_e$) is much larger than that for holes ($|\gamma_h|$). Since electrons have smaller effective mass, their states are much more sensitive to the change of QD shape or size than considerably heavier holes, which have in our calculation predominantly heavy-hole character. It follows, that $\gamma_e$ ($b = 117$) is more sensitive to variation of QD size and also its magnitude is larger than that of $|\gamma_h|$ ($b = 18$).

On the other hand, the lateral confinement does not change in the case of the variation of $h$ for fixed $d$, see Fig.~\ref{fig2}~(b). However, despite that, $\gamma$ grows slightly. First, we describe the height dependence for electrons and note that the value of $\gamma$ also depends on the effective mass, cf. Eq.~(\ref{eq2.2}). It was shown previously for InAs/GaAs QDs that the effective mass of electron and hole depends on QD height and base diameter~\cite{Zhou2009}. The effective mass of the electron decreases with increasing $h$, what also explains the increase of $\gamma_e$ ($b = 34$) in our calculations. The case of the height dependence of $|\gamma_h|$ ($b = 7$) is, however, more complex. The effective mass of heavy-holes (HH) grows with increasing $h$~\cite{Zhou2009}. On the other hand, increasing $h$ leads to admixture of $\ket{\rm LH}$ states due to larger amount of $\ket{p_z}$ Bloch waves, the contribution of which increases for higher QDs which might even consist of purely $\ket{\rm LH}$ states~\cite{Huo2014}. Moreover, $|\gamma_h|$ varies very slowly, hence, we might assume that the two aforementioned effects nearly cancel each other.

Lastly, we fix the aspect ratio of QD and change both $d$ and $h$, $h/d = 3/13$. We observe the steepest change of $\gamma_e$ ($b = 176$). Here, both the reduction of the electron effective mass and the reduction of lateral confinement contribute to the increase of $\gamma_e$. In the case of $|\gamma_h|$ ($b = 10$) we can see combination of two opposing trends, discussed before. The reduction of lateral confinement caused by increasing $d$ leads to the increase of $|\gamma_h|$ while larger $h$ slightly reduces that. This results in a slower growing trend of $|\gamma_h|$ as we can see in Fig.~\ref{fig2}~(c).

Using the SP approach we can write that $\gamma$ of SP electron--hole pair is~\cite{vanBree2012}
\begin{equation}
\gamma_{\rm{SP}} = |\gamma_e|+|\gamma_h|.
\end{equation}
The parameter $\gamma_{\rm SP}$ is mostly influenced by the electronic part of electron--hole pair SP transition which we mark as $\mathcal{X}^0$, since we omit the effect of the Coulomb interaction. As we can see from Fig.~\ref{fig2}, the presence of the direct and the exchange Coulomb interaction slightly reduces the values of $\gamma_{\rm CI}$ computed by CI. The effect of correlations influences the excitonic $\gamma_{\rm \rm CI}$ rather weakly. In Figs.~\ref{fig2}~(a)~and~(c) we observe that the deviation between $\gamma_{\rm SP}$ and $\gamma_{\rm CI}$ increases with growing size of QD. 
However, relative deviation between $\gamma_{\rm SP}$ and $\gamma_{\rm CI}$ is in the whole studied range of QD sizes rather small and, thus, we can conclude that SP approximation reasonably well describes the diamagnetic coefficient of the ground state of $\rm X^0$.

\subsection{g-factors}

We divide this section into three parts. Firstly, we discuss the electronic $g$-factor ($g_e$), we follow by the hole $g$-factor ($g_h$), and finally, the excitonic $g$-factor, computed both using SP approach ($g_{\rm SP}$) and CI ($g_{\rm CI}$), respectively, is considered. Studied dependencies are shown in the middle column of Fig.~\ref{fig2}.

\subsubsection{Electron g-factor}


\begin{figure}[htbp]
	\centering
		\includegraphics[width=87mm]{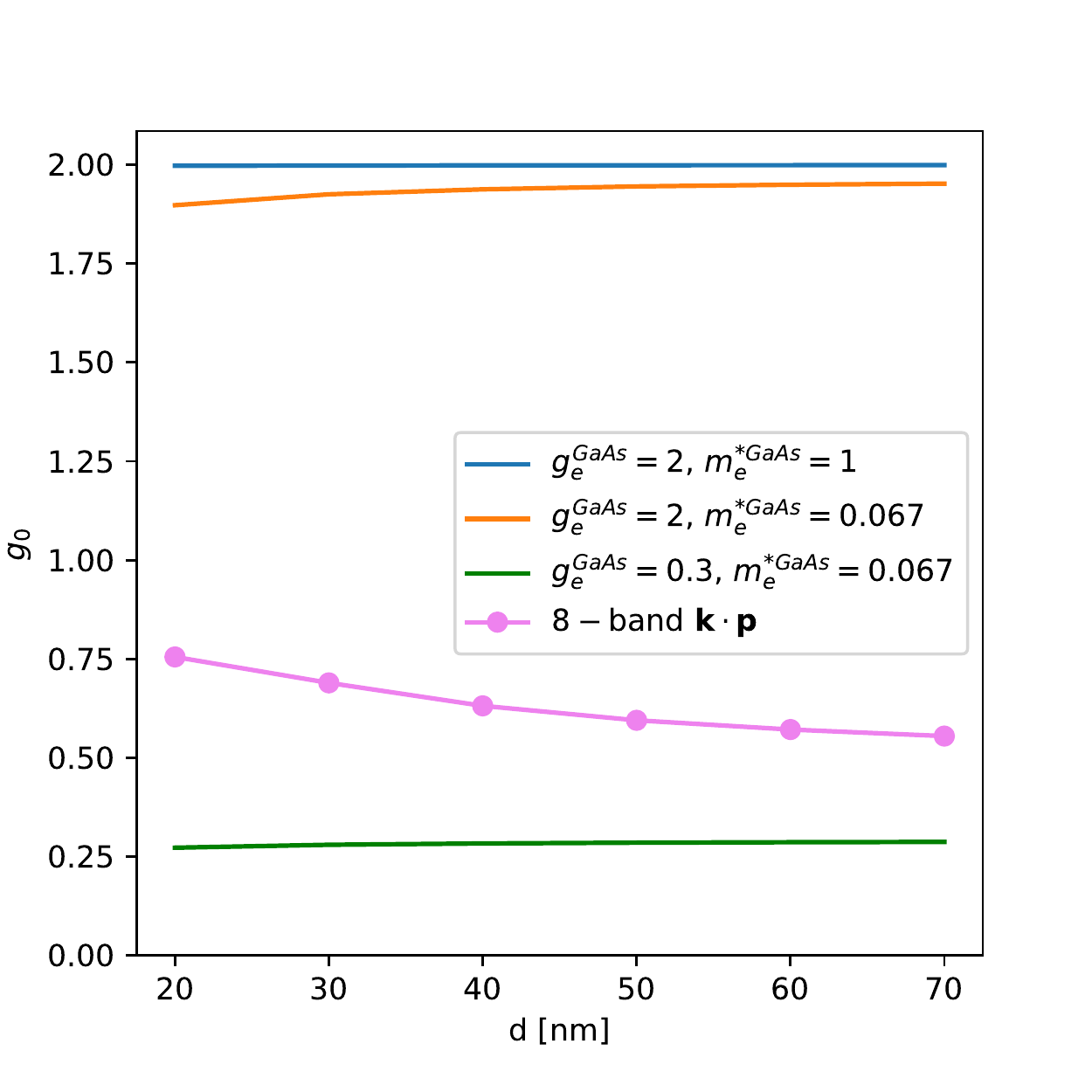}
	\caption{Comparison of dependencies of the electronic $g_e$ computed by single-band $\textbf{k}\cdot\textbf{p}$ method for various values of the bulk $g_e^{\rm GaAs}$ and electron effective mass $m_e^{*\rm GaAs}$. The data are shown for $g_e^{\rm GaAs}=2$ and $m_e^{*\rm GaAs}=1$ (blue curve), $g_e^{\rm GaAs}=2$ and $m_e^{*\rm GaAs}=0.067$ (orange curve), and $g_e^{\rm GaAs}=0.3$ and $m_e^{*\rm GaAs}=0.067$ (green curve), respectively. To ease comparison, we show also SP electron $g_e$ computed using eight-band $\textbf{k}\cdot\textbf{p}$ (violet curve and symbols), taken from Fig.~\ref{fig2}~(a).}
	\label{figElgFact}
\end{figure}

\begin{figure}[htbp]
	\centering
		\includegraphics[width=87mm]{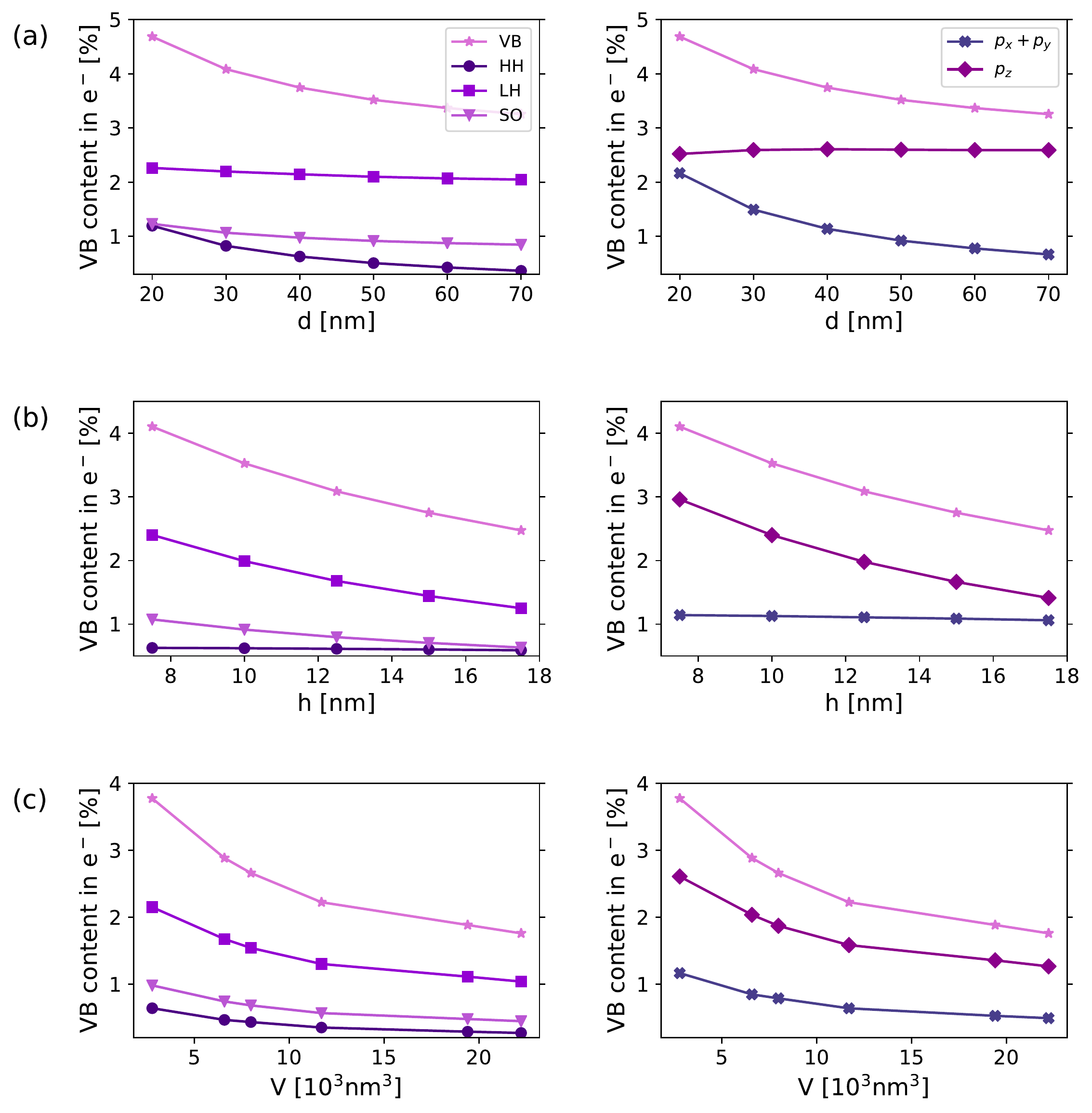}
	\caption{Dependencies of the amount of $\rm{\ket{HH}}$ (circles), $\rm{\ket{LH}}$ (squares), and $\rm{\ket{SO}}$ (triangles) components, that for $\ket{p_x}$ and $\ket{p_y}$ (cross), and $\ket{p_z}$ (diamonds) Bloch waves, and the total VB contribution,~i.e.~$\mathcal{N}(\rm{HH}) + \mathcal{N}(\rm{LH}) + \mathcal{N}(\rm{SO})$ (stars) in the electron ground state for $B_z = 0\,\rm{T}$, respectively. The calculations are shown as a function of QD size for (a) fixed height of $h = 9\,\rm nm$, (b) fixed base diameter of $d = 40\,\rm nm$, and (c) fixed aspect ratio of $h/d = 3/13$, where $V$ is the volume of QD. Note, that $p_x+p_y$ in the inset of (a) indicates that contents of $\ket{p_x}$ and $\ket{p_y}$ states were added, not the actual Bloch waves.} 
	\label{fig3}
\end{figure}

The value of $g_e$ is $+2$ in the limit of infinite confinement, which is the result of the quenching of the angular momentum~\cite{Pryor2006}. In bulk semiconductors that value is reduced in $\textbf{k}\cdot\textbf{p}$ method due to spin-orbit coupling $\Delta_{\rm SO}$ and the electron effective mass $m^*_{e}$, the magnitude of the reduction of $g_e$ being due to $\Delta_{\rm SO}$ and $m^*_{e}$, respectively~\cite{t_Andlauer}. Thus,~e.g., in GaAs the value of $g_e\approx 0.3$ is attained~\cite{Oestreich1996}. The aforementioned decomposition is illustrated in Fig.~\ref{figElgFact} where we plot that for $d$ variation of our GaAs QDs. Clearly, the dominant reduction of $g_e$ from 2 is caused by the interaction of electron with the crystal lattice potential, characterized in the single-band $\textbf{k}\cdot\textbf{p}$ by bulk $g_e^{\rm GaAs}$. Furthermore, the size dependence of $g_e$ observed in Fig.~\ref{fig2} is caused by the admixture of Bloch states from the valence bands (VB),~i.e,~heavy-hole ($\rm \ket{HH}$), light-hole ($\rm \ket{LH}$), and split-off ($\rm \ket{SO}$) Bloch states, into the ground state of electron and we proceed by discussing the reasons for that (see also Ref.~\cite{vanBree2012}). In the eight-band $\textbf{k}\cdot\textbf{p}$ calculations we express each quantum state as a superposition of $\rm \ket{S}$, $\rm \ket{HH}$, $\rm \ket{LH}$, and $\rm \ket{SO}$ bulk Bloch components. The wave functions consist of the Bloch and the envelope parts with (total) angular orbital momenta $\mathbf{J}$ and $\mathbf{L}_E$, respectively, which are coupled due to the spin-orbit interaction. Since the $\rm \ket{S}$ Bloch component has $\mathbf{L}_E = 0$, it does not influence $g_e$ at all. However, the Bloch functions in VB have envelope angular momenta $\mathbf{L}_E = 1$ and, thus, $L_{Ez} = \{0,\pm 1\}$. Hence, the further deviation of $g_e$ from the value of $+2$ in the case of multi-band $\textbf{k}\cdot\textbf{p}$ is caused by the admixture of the VB Bloch functions, which have $L_{Ez} = \pm 1$, into electron envelopes. Since the coupling between the states from conduction band (CB) and states from VB with $L_{Ez} = \pm 1$ is proportional to the crystal momentum vector components $k_x$ and $k_y$~\cite{vanBree2012}, we get the content of states having $L_{Ez} = \pm 1$ by taking into account $\ket{p_x}$ and $\ket{p_y}$ Bloch waves, mixed in the ground state of the electron. Furthermore, the coupling with the components which have $L_{Ez} = 0$ is proportional to $k_z$~\cite{vanBree2012}, hence, we get these states as $\ket{p_z}$ Bloch waves. 

In contrast with $g_e$ for cylindrical InAs/InP QDs studied in Ref.~\cite{vanBree2012}, $g$-factors of our QDs depend on their size rather weakly. Bulk GaAs has band-gap of $E_g = 1.43\,\rm eV$~\cite{ioffe} (for temperature of $300\,\rm{K}$), which is nearly four times larger than that for bulk InAs, resulting in weaker mixing of CB and VB states in the case of GaAs QDs. The size dependence of the sum of amounts of $\ket{p_x}$ and $\ket{p_y}$ Bloch waves (states with $L_{Ez} = \pm 1$) in electron ground state for $B_z = 0\,\rm{T}$ is shown in Fig.~\ref{fig3} (right column). Note, that for the sake of completeness, we also show in Fig.~\ref{fig3} the size dependencies of $\rm \ket{HH}$, $\rm \ket{LH}$, and $\rm \ket{SO}$ components (left column). In all cases of the studied size variations, the content of states from VB decreases with increasing size by $\sim 2\,\%$. The sum of all $L_{Ez} = \pm 1$ components decreases with increasing $d$ \{Fig.~\ref{fig3}~(a)\} by $\sim 10$-times smaller rate compared to that for InAs/InP QDs~\cite{vanBree2012}. Interestingly, in the case of the variation of $h$ \{Fig.~\ref{fig3}~(b)\} we observe even $\sim 24$-times reduced rate. Since the admixture of the VB components, which affect $g_e$, depends on the structural properties of QDs rather weakly we do not observe strong size dependence of $g_e$. For the completeness we also show the size dependence for the studied parameters in the case of fixed aspect ratio \{Fig.~\ref{fig3}~(c)\}, even though we cannot compare to any similar study for InAs/InP QDs.

\subsubsection{Hole g-factor}

\begin{figure}[htbp]
	\centering
		\includegraphics[width=85mm]{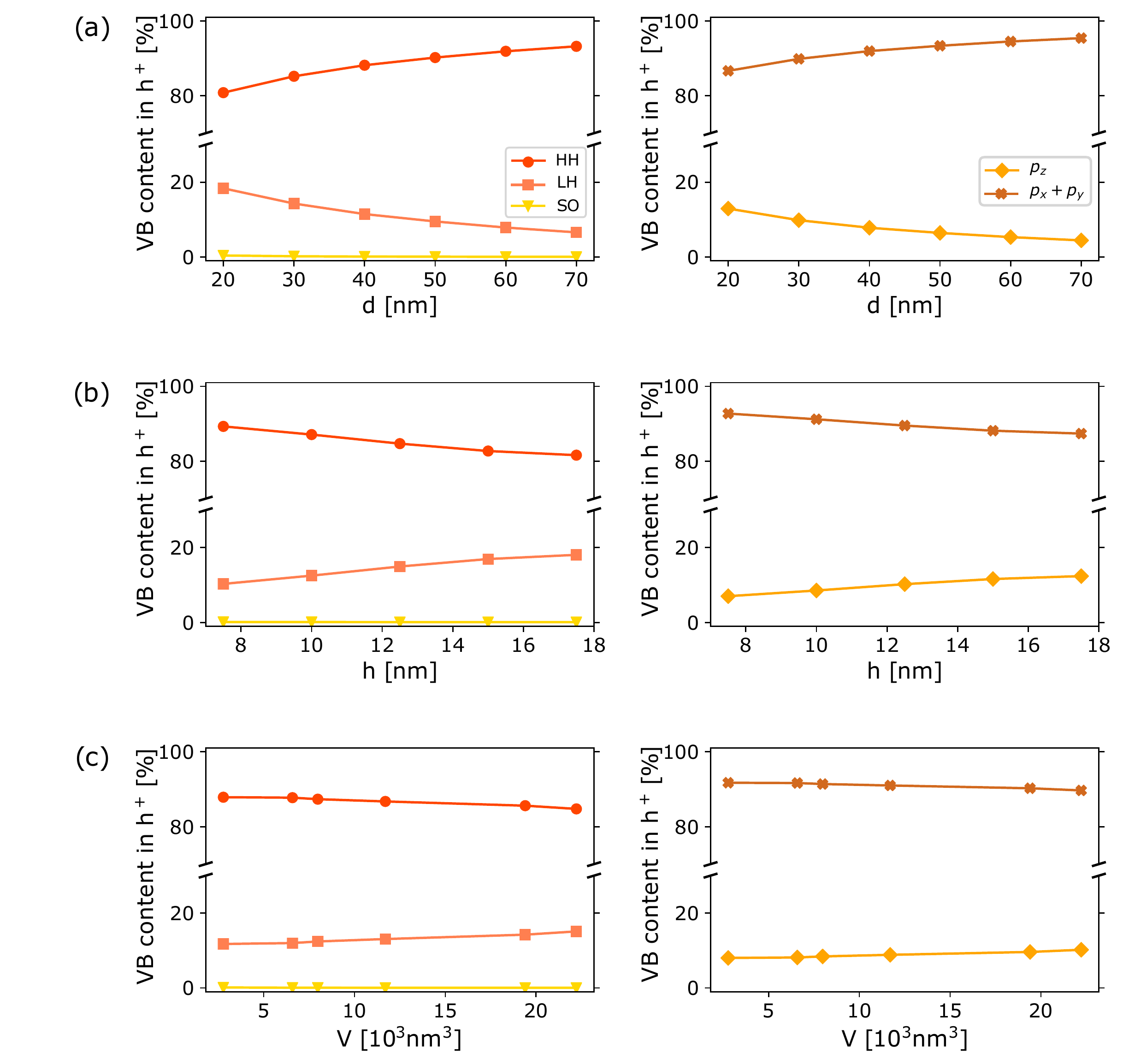}
	\caption{Dependencies of $\rm{\ket{HH}}$ (circles), $\rm{\ket{LH}}$ (squares), and $\rm{\ket{SO}}$ (triangles) components, that for $\ket{p_x}$ and $\ket{p_y}$ (cross), and $\ket{p_z}$ (diamonds) Bloch waves in hole ground state for $B_z = 0\,\rm{T}$, respectively. The calculations are shown as a function of QD size for (a) fixed height of $h = 9\,\rm nm$, (b) fixed base diameter of $d = 40\,\rm nm$, and (c) fixed aspect ratio of $h/d = 3/13$, where $V$ is the volume of QD. Note, that $p_x+p_y$ in the inset of (a) indicates that contents of $\ket{p_x}$ and $\ket{p_y}$ states were added, not the actual Bloch waves.} 
	\label{fig4}
\end{figure}

In the case of $g_h$ we observe a rather strong size dependence of that. Since in the case of holes the VB band mixing plays much more prominent role in $g$-factor, we do not apply the approach which we discussed above for electrons~\cite{vanBree2012}. However, there exists a connection between the value of $g_h$ and HH-LH mixing in the hole ground state~\cite{Ares2011}. The approach introduced in Ref.~\cite{Ares2011} utilizes 2D effective model and can be expressed as~\cite{Ares2011,Bayer2002}
\begin{equation}
    g_h = 6\kappa + \frac{27}{2}q - 2\gamma_{lh},
\label{eq2.7}
\end{equation}
where $\kappa$ and $q$ are the Luttinger parameters in the standard notation~\cite{Luttinger1956}, which are for GaAs summarized in Table~\ref{tab2.1}, and $\gamma_{lh}$ is given by the overlap of $\rm \ket{HH}$ and $\rm \ket{LH}$ states of the hole. It follows from the Eq.~\eqref{eq2.7} that $g_h$ strongly depends on the bulk properties of the QD material and, furthermore, that the content of $\rm \ket{LH}$ states in the hole reduces the magnitude of $g_h$, since $\gamma_{lh}$ attains positive values~\cite{Ares2011}.
The validity condition for the discussed 2D model is that $h/d \ll 1$, which is fulfilled for our QDs. Furthermore, it is assumed in the model that the mixing of $\rm \ket{SO}$ states can be neglected since
\begin{equation}
    E \ll \Delta_{\rm HH-LH} \ll \Delta_{\rm SO},
    \label{eq2.8}
\end{equation}
where $E$ is the energy measured far-away from the edge of the top-most subband, $\Delta_{\rm SO}$ is the spin-orbit energy, and $\Delta_{\rm HH-LH}$ is the energy splitting of $\rm \ket{HH}$ and $\rm \ket{LH}$ states, caused by the confinement and/or the biaxial strain. As one can see in Fig.~\ref{fig4}, the contribution of $\rm \ket{SO}$ Bloch waves in the hole ground state is minuscule in comparison with the other two components. The reason of small admixture of $\rm \ket{SO}$ Bloch waves into the hole ground state is the large bulk value of $\Delta_{\rm SO}$ (for GaAs $\Delta_{\rm SO} = 0.34\rm\,eV$ for the temperature of $300\rm\,K$~\cite{Vurgaftman2001}). On the other hand, since GaAs and $\rm Al_{0.4}Ga_{0.6}As$ are nearly lattice matched, the biaxial strain in QD is rather small, which results in small value of $\Delta_{\rm HH-LH}<1\,{\rm meV}$~\cite{Huo:NatPhys}. Hence, the condition in Eq.~\eqref{eq2.8} is fulfilled for our dots.

In the case of our calculations, the HH-LH coupling is caused by the variation of QD size, see right column of Fig.~\ref{fig2}. The admixture of $\rm \ket{LH}$ Bloch waves to the hole state depends on the content of $\ket{p_z}$ Bloch waves. That content increases with $h$. On the other hand, when $d$ is increased, the content of $\ket{p_x}$ and $\ket{p_y}$ Bloch waves, which have predominantly HH character~\cite{Birner2011}, increases as well. Consequently, the reduction of the amount of $\rm \ket{LH}$ is observed. 

The aforementioned model in Eq.~\eqref{eq2.7} provides a reasonably good qualitative prediction of the trend of $g_h$ for increasing $h$. As expected, with increasing $h$ the contribution of $\rm \ket{LH}$ Bloch waves grows \{see Fig.~\ref{fig4}(b)\}, which leads to the reduction of $g_h$ \{see Fig.~\ref{fig2}~(b)\}. For $h = 17.5\rm\,nm$ $g_h$ changes its sign. Since mathematically $g_0$ corresponds to the slope of $E_{\uparrow}(B_z) - E_{\downarrow}(B_z)$, the change of the sign of $g_h$ indicates that the order of the levels in the corresponding Kramers doublet reverses in energy, which we also confirmed by inspecting the computed states. If it would be experimentally meaningful to give the calculations for the temperature of $0\rm\,K$, the magnetization and the magnetic susceptibility of the system in a certain state would lead to the change of the sign of the susceptibility with increasing $h$~\cite{Jahan2019,Klenovsky2019}. However, since we want our results to be reproducible by experiment, we strictly performed our calculation for finite temperatures. 

\begin{table}[h!]
\begin{center}
\caption{Luttinger parameters for GaAs and AlAs~\cite{Sodagar2009,Lawaetz1971, Kiselev2001}.} 
\begin{tabular}{|c|c|c|c|c|c|c|}
\hline
     & $\gamma_1$ & $\gamma_2$ & $\gamma_3$  & $\kappa$ & $q$ \\ \hline
GaAs &      $6.98$      &     $2.06$       &      $2.93$          & $1.2$ &   $0.04$  \\ \hline
AlAs &       $3.76$     &       $0.82$     &      $1.42$          & $0.12$ &  $0.03$  \\
\hline
\end{tabular}
\label{tab2.1}
\end{center}
\end{table}

\begin{figure*}[htbp]
	\centering
		\includegraphics[width=160mm]{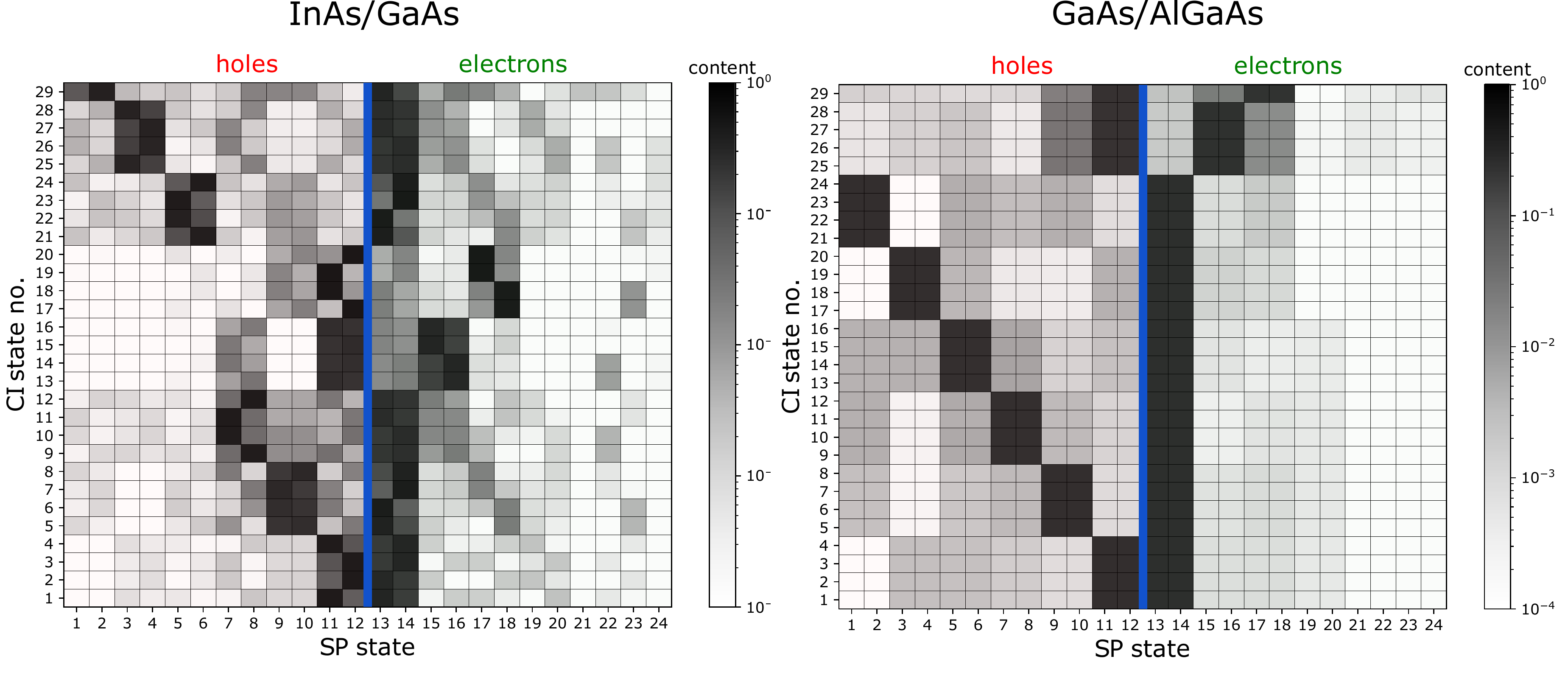}
	\caption{The contribution of SP states, used as the basis of the CI calculations, in X$^0$ without the application of external fields for InAs/GaAs pyramidal-shaped QD with base width of $20\,\rm nm$ and height of $3\,\rm nm$ (left) and GaAs/AlGaAs QD shown in Fig.~\ref{fig1}~(a) with $d = 68\,\rm nm$ and $h = 16\,\rm nm$ (right). The CI calculations were performed with the basis of 12 SP electron (green) and 12 SP hole (red) states and were carried out exactly the in the same manner. Note that the CI states are ordered from 1 to 29 according to their energy from lowest to highest. The contents of SP states are shown on logarithmic scale.} 
	\label{fig9}
\end{figure*}

Interestingly, even though the content of $\rm \ket{LH}$ Bloch waves decreases with growing $d$ \{see Fig.~\ref{fig4}~(a)\}, we observe a slow decrease of $g_h$ as well. Similar trend was previously observed for InAs pyramidal QDs in Ref.~\cite{Nakaoka2004}. We assume that also other effect causes the reduction of $g_h$, apart of those previously discussed. Since the decrease of $g_h$ is the weakest of all the discussed cases, we assume that the effect which reduces $g_h$ is similarly strong, when $d$ grows, as the decrease of the content of $\rm \ket{LH}$ components. The parameter $g_h$ also depends on the bulk material parameters, see Eq.~\eqref{eq2.7}. As the lateral quantum confinement becomes weaker with increasing $d$, the hole wave function moves towards the top of the QD. If hole wave function would partially leak out of the QD material, the change in $g_h$ might have been affected also by the properties of the surrounding $\rm Al_{0.4}Ga_{0.6}As$. In Table~\ref{tab2.1} we summarize the values of $\kappa$ for GaAs and AlAs. Using the linear interpolation we can estimate $\kappa$ of $\rm Al_{0.4}Ga_{0.6}As$ as $\kappa = 1.08$. While the smaller value of $\kappa$ would lead to the reduction of $g_h$, an inspection of the probability density of hole ground state shows leakage of hole out of QD only for QDs with $d > 30\rm\,nm$. Since, we observe the reduction of $g_h$ also for smaller $d$ we conclude that this effect is not strong enough to cause the decreasing trend of $g_h$.

\begin{figure}[htbp]
	\centering
		\includegraphics[width=85mm]{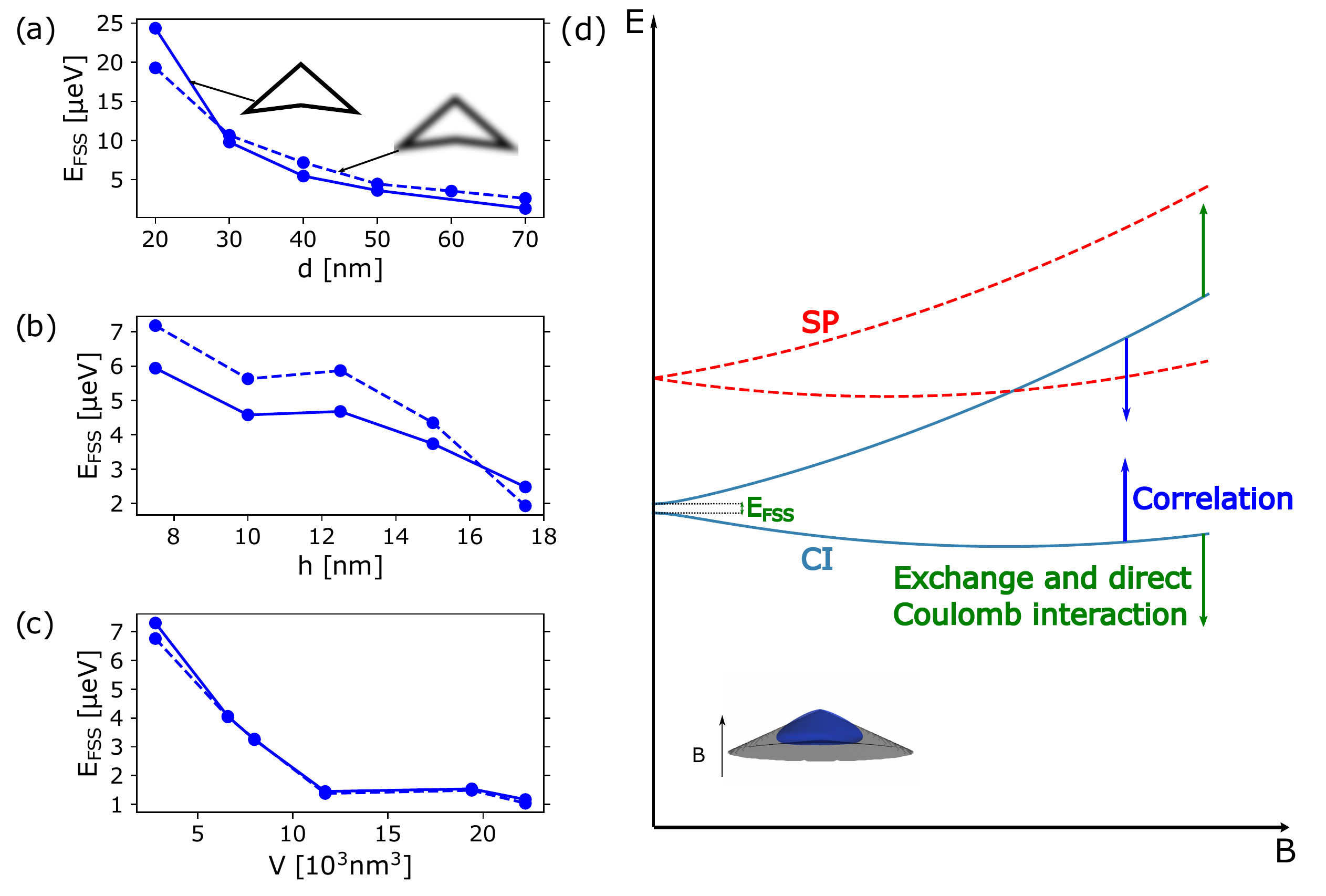}
	\caption{Dependencies of $E_{\rm FSS}$ for QD structure with sharp interface between GaAs and $\rm Al_{0.4}Ga_{0.6}As$ (solid curve) and for graded composition from GaAs to $\rm Al_{0.4}Ga_{0.6}As$ in the vicinity of the QD-bulk interface within 1~nm around that (broken curve), see also sketch in panel (a). The CI calculations with $12\times12$ basis are shown as a function of QD size for (a) fixed height of $h = 9\,\rm nm$, (b) fixed base diameter of $d = 40\,\rm nm$, and (c) fixed aspect ratio of $h/d = 3/13$, where $V$ is the volume of QD. (d) Sketch of the effect of the direct and the exchange Coulomb interactions, respectively, and the correlation on the energy splitting of levels due to $B_z$.} 
	\label{figFSS}
\end{figure}

We observe strong dependence of $g_h$ on $V$ of QD for fixed QD aspect ratio. That effect is caused by the admixture of $\rm \ket{LH}$ Bloch waves,~see Fig.~\ref{fig4}~(c), similarly as in the previous case. Also $E_{\rm{X}^0}$ for which $g_h$ crosses zero is similar for the cases of fixed aspect ratio and fixed $h$, see Fig.~\ref{fig2}.

\subsubsection{Excitonic g-factor}

We further fitted the $B_z$ dependence of the difference of the ground state Kramers doublet energies of $\mathcal{X}^0$, by Eq.~(\ref{eq2.2}) and obtained the slope of that which we mark as $g_{\rm SP}$, see middle column of Fig.~\ref{fig2}. From the SP approach we find that $g_{\rm SP} = g_e + g_h$ of the bright state, similarly as in Refs.~\cite{Bayer2002,Witek2011}.
Resulting from that and already discussed properties of $g_h$ and $g_e$, it follows that the trend of $g_{\rm SP}$ is dominated by the hole part of $\mathcal{X}^0$. Since the value of $g_e$ remains nearly constant with QD size change, it only causes the increase of the mean value of $g$-factor, leading to the zero-crossing of $g_{\rm SP}$ for smaller emission energies. To see how the HH-LH coupling affects $g_{\rm SP}$, we show in the last column of Fig.~\ref{fig2} the content of $\rm \ket{LH}$ Bloch wave in the exciton. The calculations were done for $\mathcal{X}^0$ and for X$^0$ computed by CI with basis of $2 \times 2$ and $12 \times 12$ electron and hole SP states, respectively (see also Appendix II). Clearly, the HH-LH coupling is not influenced by the effect of correlation, however, it is weakly affected by the direct and the exchange Coulomb interactions. The connection between band mixing and $g_{\rm SP}$ is the same as that already discussed for holes.

The inclusion of the direct and the exchange Coulomb interaction ($2 \times 2$ basis) causes the overall increase of the $g$-factor. Note, that the difference between $g_{\rm SP}$ and $g_{\rm CI}$ computed in $2 \times 2$ basis grows with the size of QD. As a result of the exchange and the direct Coulomb interaction, $g_{\rm CI}$ does not cross zero for the range of considered sizes, see panels (b) and (c) of the middle column of Fig.~\ref{fig2}.

On the other hand, the effect of correlation causes the reduction of $g_{\rm CI}$. Hence, there are multi-particle effects which affect the Zeeman splitting and $g$-factor in opposite ways. The direct and exchange Coulomb interaction amplifies the Zeeman splitting, increasing $g_{\rm CI}$, yet the correlations cause the reduction of that. The effect of multi-particle effects on the Zeeman splitting is sketched in Fig.~\ref{figFSS}~(d). 
As we can see in Fig.~\ref{figFSS}~(d) and Eq.~\eqref{eq2.3} $E_{\rm FSS}$ is the initial value of the Zeeman splitting. The size dependence of $E_{\rm FSS}$ is shown in Fig.~\ref{figFSS}~(a)--(c). As already discussed,~e.g., in Ref.~\cite{Huo2014}, $E_{\rm FSS}$ decreases with increasing size of QDs. Since the envelope--function approximation cannot model the effect of the atom disorder, it cannot describe effects such as the interface roughness, which occurs in real structures and might affect quantities such as $E_{\rm FSS}$. However, we simulated that by smearing the alloy composition of QD in the vicinity of GaAs/Al$_{0.4}$Ga$_{0.6}$As interface. That allows us to model the effect of the spatially mean interface roughness. The size dependence of $E_{\rm FSS}$ calculated in $12\times 12$ basis of such structure is shown in Fig.~\ref{figFSS}~(a)--(c) by broken curves. As can be seen, the smeared alloy composition changes the values of $E_{\rm FSS}$ in most cases within~$\pm 1\,\mu\textrm{eV}$,~i.e., corresponding to the expected error of our CI calculations. Hence, we can conclude that the influence of the mean interface roughness on $E_{\rm FSS}$ is negligible in our calculations.

Finally, we note that interface roughness or alloy disorder can be included using atomistic methods like,~e.g., pseudopotentials or tight-binding which are not the scope of this work. Nevertheless, we compare our values of FSS in Fig.~\ref{figFSS} with that obtained for similar structures by (i) the combination of pseudopotentials and CI in Ref.~\cite{Luo2012} and (ii) experimentally measured statistics in Ref.~\cite{Huo:APL2013}, being $5\pm1.4\,\mu{\rm eV}$ and $3.9\pm1.8\,\mu{\rm eV}$ in the former and latter cases, respectively,~i.e., similar as our values.


Moreover, we stress that the values of FSS in Fig.~\ref{figFSS} as well those for $g$-factor or $\gamma$ do not correspond to any particular experimental QD geometry or composition \{see also Fig~\ref{fig1}~(b)\}. Thus, in realistic QDs the values of the aforementioned parameters might be slightly different.

Note, that for increasing $h$ of QD the difference between CI calculation without and with the effect of correlation grows, see panel (b) of the middle column of Fig.~\ref{fig2}. Surprisingly, for the calculations where only $d$ is varied, the deviation is not systematic \{panel (a) of the middle column of Fig.~\ref{fig2}\}, thus, it seems that the lateral size of QD does not influence the effect of correlation on $g_{CI}$. This observation is unexpected since the trend of increasing difference between calculations with and without the effect of correlation is stronger for varying $V$ with fixed aspect, what is the consequence of the faster decrease of the quantum confinement due to increasing of both $h$ and $d$.

To further visualize the importance of the correlation for the SP-state-resolved description of X$^0$ we show in Fig.~\ref{fig9} the content of hole and electron SP states in those of CI for a prototypical pyramidal-shaped InAs/GaAs QD with the base width of $20\,\rm nm$ and height of $3\,\rm nm$ and GaAs/Al$_{0.4}$Ga$_{0.6}$As dot, shown in Fig.~\ref{fig1}, with the base diameter $68\,\rm nm$ and height $16\,\rm nm$. Here, the level of darkness identifies the content of each SP state,~i.e., the darker the larger content. On the vertical axis the number of the CI eigenstate is shown. The states are ordered by energy,~e.g., $1,2$ is the ground state doublet. The numbers on the horizontal axis represent the numbers of the SP electron (green) or hole (red) states computed by Nextnano software~\cite{Birner2007}. Here, numbers $11,12$ $\{13,14\}$ mark hole $\{$electron$\}$ ground states. In the former case (InAs/GaAs QD), the first four CI ground states of X$^0$ (CI states 1--4 in Fig.~\ref{fig9}) clearly show the dominant contribution of particular hole or electron SP state to a given CI state. That is smeared out for GaAs/Al$_{0.4}$Ga$_{0.6}$As QD. In the latter case, the correlation causes via the exchange interaction the mixing of the almost energy degenerate hole SP states, competing, thus, with the Zeeman interaction, as sketched in Fig.~\ref{figFSS}~(d). The details about the evaluation of the contents of SP states in the CI states are discussed in Appendix II.

Moreover, we note that we performed the convergence tests of our CI computations and the effect of correlation by comparing results of $E_{\rm FSS}$ of the largest dot (and thus largest effect of correlation) for CI bases of $2\times 2$, $6\times 6$, and $12\times 12$, respectively. We observed an energy difference for $E_{\rm FSS}$ between $6\times 6$ and $12\times 12$ bases, respectively, of $\approx 0.05\,\mu{\rm eV}$ being below the numerical resolution of the method and we can, thus, safely regard the results for $12\times 12$ basis as converged.




\subsection{Comparison with the experimental data}\label{sec2.3.3}

\begin{figure}[htbp]
	\centering
		\includegraphics[width=85mm]{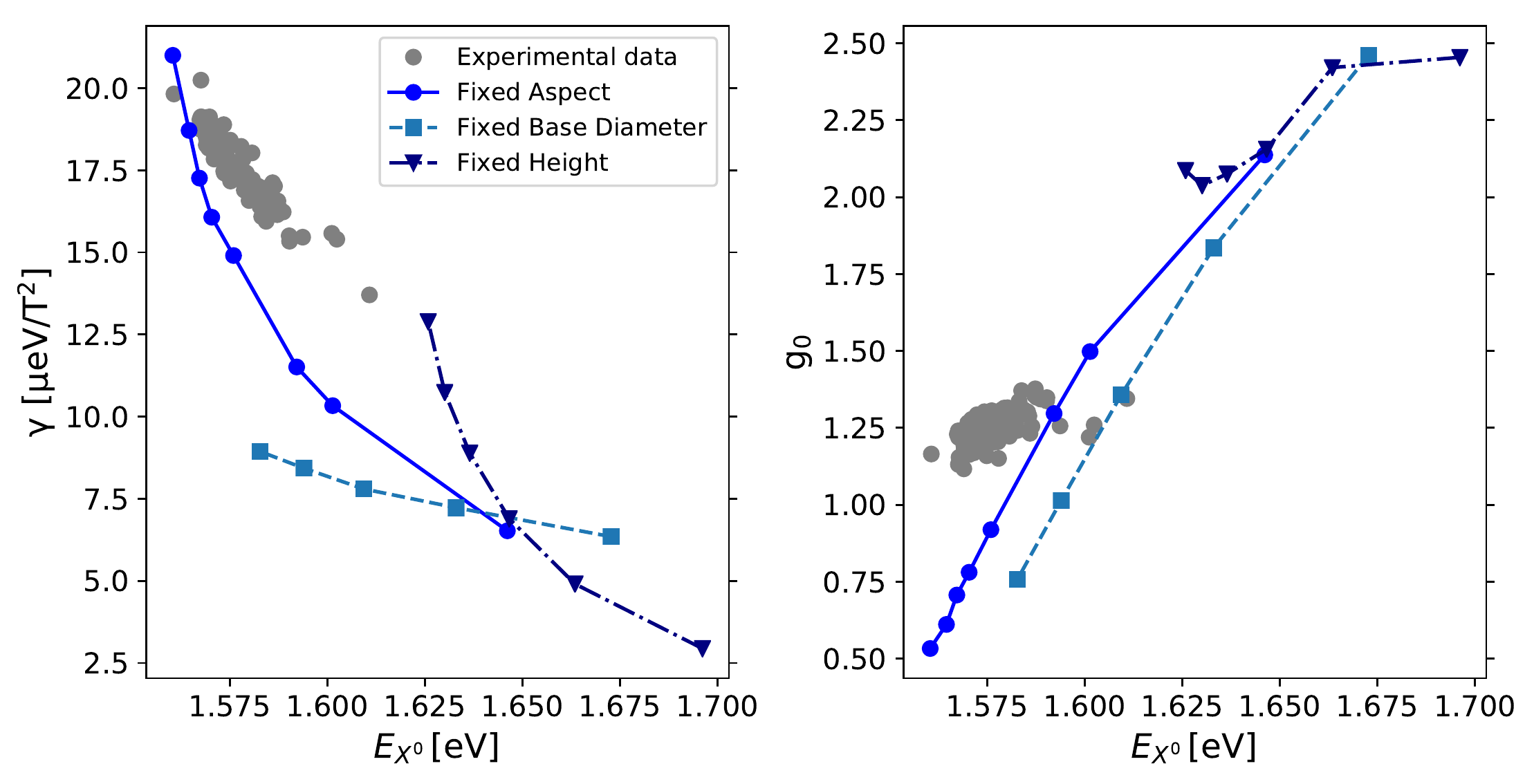}
	\caption{Dependencies of X$^0$ ground state magneto-optical properties on the emission energy $E_{\rm{X}^0}$ (blue). The experimental measurements were reprinted from Ref.~\cite{Lobl2019} (grey).} 
	\label{fig6}
\end{figure}

Finally, we compare our calculations with the measurements performed by M.~C.~L\"{o}bl and colleagues~\cite{Lobl2019}, see Fig.~\ref{fig6}. The samples were grown by MBE on the $(001)$ substrate. The measured QDs were, as well as those discussed in the previous sections, cone shaped and had also the same composition,~i.e. pure GaAs QDs embedded in Al$_{0.4}$Ga$_{0.6}$As. The authors further assumed that all the Al droplets which etched the substrate had the same aspect ratio. Since the relation between Al droplet height $h_d$ and QD height $h$ is given by the phenomenological relation $h \propto h_d^{\beta}$~\cite{Atkinson2012}, we assume that the aspect ratio of measured QDs was the same for all QDs and, thus, we compare with that our calculations where the aspect ratio of QDs is fixed as well. However, we show in our comparison all the size dependencies in order to indicate whether the aforementioned assumption is correct. For further details about the growth and measurements we refer the reader to Ref.~\cite{Lobl2019}.

For the comparison we use the magneto-optical properties of multi-particle X$^0$ where the effect of correlation is included. As we can see, the calculated trends reasonably fit the measurements. However, we observe larger disagreement with the size dependence of the measured magneto-optical properties for larger emission energies $E_{\rm{X}^0}$,~i.e., smaller QDs. The deviation between theory and experiment might be,~e.g., attributed to the fact that we did not optimize the QD shape to fit the measurements more precisely and, thus, the aspect ratio or shape of the base of the calculated QDs can be slightly different than that for experiment. Moreover, spatial variation of the chemical composition inside QD might also affect the slopes of calculated trends.

Let us first compare theoretically and experimentally obtained $\gamma$.
As we can see, almost all the calculated dependencies have similar non-linear trends. They decrease fast for smaller $E_{\rm X^0}$ and from certain value the decreases is approximately linear. The calculated dependencies resemble a combination of two linear trends with different slopes or a single exponential decrease. The linear trend shows also $\gamma$ determined from the experiment. Generally, by comparing the steepness of the experimentally determined data and calculations, we deduce that measured QDs were slightly larger than those calculated.

In the case of $g$-factors we observe larger differences between calculations and experiment. Here, the slopes of calculated dependencies are unfortunately significantly larger than that of the measured data for smaller values of $E_{\rm \rm{X}^0}$. However, we note that a more favorable correspondence of the theoretical slope with that of the experiment might be observed for variation of $d$ with fixed $h$ and at the same time smaller values of $E_{\rm \rm{X}^0}$. However, to match the experiment, one would clearly need also the dot to have larger $V$.

\section{Size dependence of magneto-optical properties of trion states in magnetic field}

\begin{figure*}[htbp]
	\centering
		\includegraphics[width=170mm]{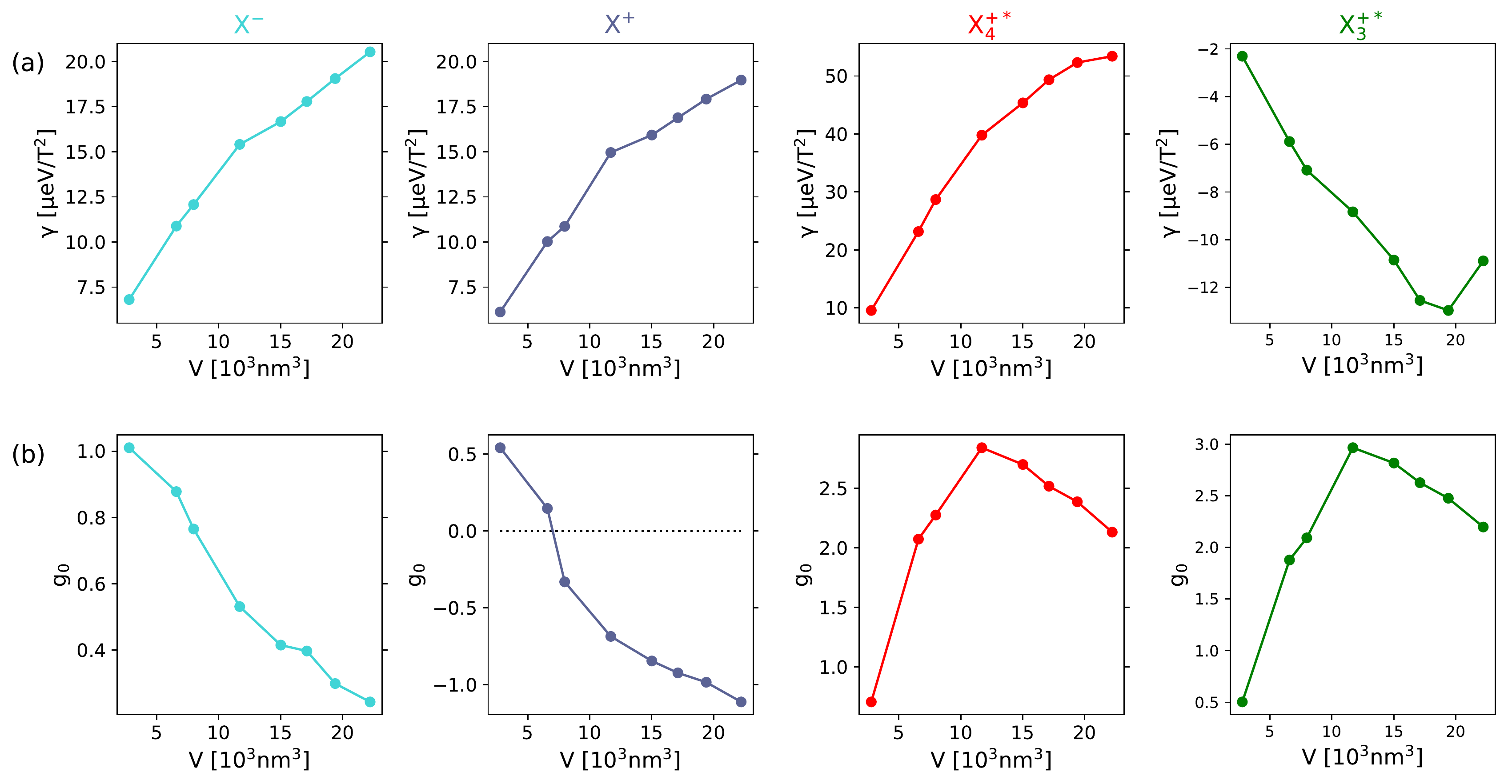}
	\caption{Size dependence of (a) $\gamma$ and (b) $g$-factor of X$^-$ (light blue), X$^+$ (blue), X$_4^{+*}$ (red), and X$_3^{+*}$ (green), respectively.}
	\label{figB.1}
\end{figure*}

In this section we expand our analysis of the size dependence of magneto-optical properties to incorporate also the positively charged ground X$^+$ and excited X$_3^{+*}$ and X$_4^{+*}$ trions, respectively, and that for excitons with surplus electron,~i.e., X$^-$. The calculations are performed for QD shown in Fig.~\ref{fig1} with fixed aspect ratio of $h/d = 3/13$, similarly as that in section~\ref{sec2.3}. The field $B$ is again applied in the $[001]$ growth direction,~i.e.~$B_z$, and the energies of the excitonic complexes are calculated by CI with the basis of $12$ SP electron and $12$ SP hole states. Studied size dependencies of magneto-optical properties are shown in Fig.~\ref{figB.1}.

The particular choice of X$_3^{+*}$ and X$_4^{+*}$ is motivated by their experimental observation in Ref.~\cite{Huber2019} and the larger contribution in the respective CI complexes which we show along with the contribution of SP states in CI complexes of X$^-$ in Appendix~III.

As expected, $\gamma$ of X$^-$ and X$^+$ have very similar trend as that for X$^0$ \{cf. Fig.~\ref{fig2}~(c)\}. To see the differences among $\gamma$ of the excitonic complexes we discuss again the parameter $b$ defined in Eq.~\eqref{eq:ParameterB}. Due to the dominant content of holes, which have larger effective mass than electron, we observe the smallest value of $b$ for X$^+$ ($b = 150$). Interestingly, $\gamma$ grows slightly more for X$^0$ ($b = 168$) than for X$^-$ ($b = 160$). That results from previous investigation of size dependence of $\gamma$ for X$^0$, since direct and exchange Coulomb interaction reduce the values of $\gamma$ for larger QDs, what leads to the reduction of the parameter $b$. Moreover, in the case of X$^-$ \{X$^+$\} the Coulomb interaction between the electron and the hole is doubled compared to X$^0$ and also that between electrons \{holes\} is included what also contributes to the reduction of $b$.

Due to the coupling of singlet (X$_4^{+*}$) and triplet (X$_3^{+*}$) states of the excited trions we observe the anomalous diamagnetic shift of X$_3^{+*}$ transition~\cite{Huber2019}. The parameter $\gamma$ of X$_3^{+*}$ is negative for all considered QDs and its absolute value increases with size until it reaches the minimum for $V \approx 20\cdot 10^{3}\,\rm nm^3$. On the other hand, $\gamma$ of X$_4^{+*}$ grows monotonically with increasing $V$. Interestingly, we observe smaller absolute values of $\gamma(\rm X_3^{+*}$) and $\gamma(\rm X_4^{+*}$) for QDs with $h = 17\,\rm nm$ and $d = 75\,\rm nm$ (or $h = 18\,\rm nm$ and $d = 78\,\rm nm$) considered in this section, than for QDs with $h = 15\,\rm nm$ and $d = 70\,\rm nm$,~see Tab.~2 in Ref.~\cite{Huber2019}. We assume that this might be caused by different aspect ratio of QDs under consideration.
\begin{figure}[htbp]
	\centering
		\includegraphics[width=60mm]{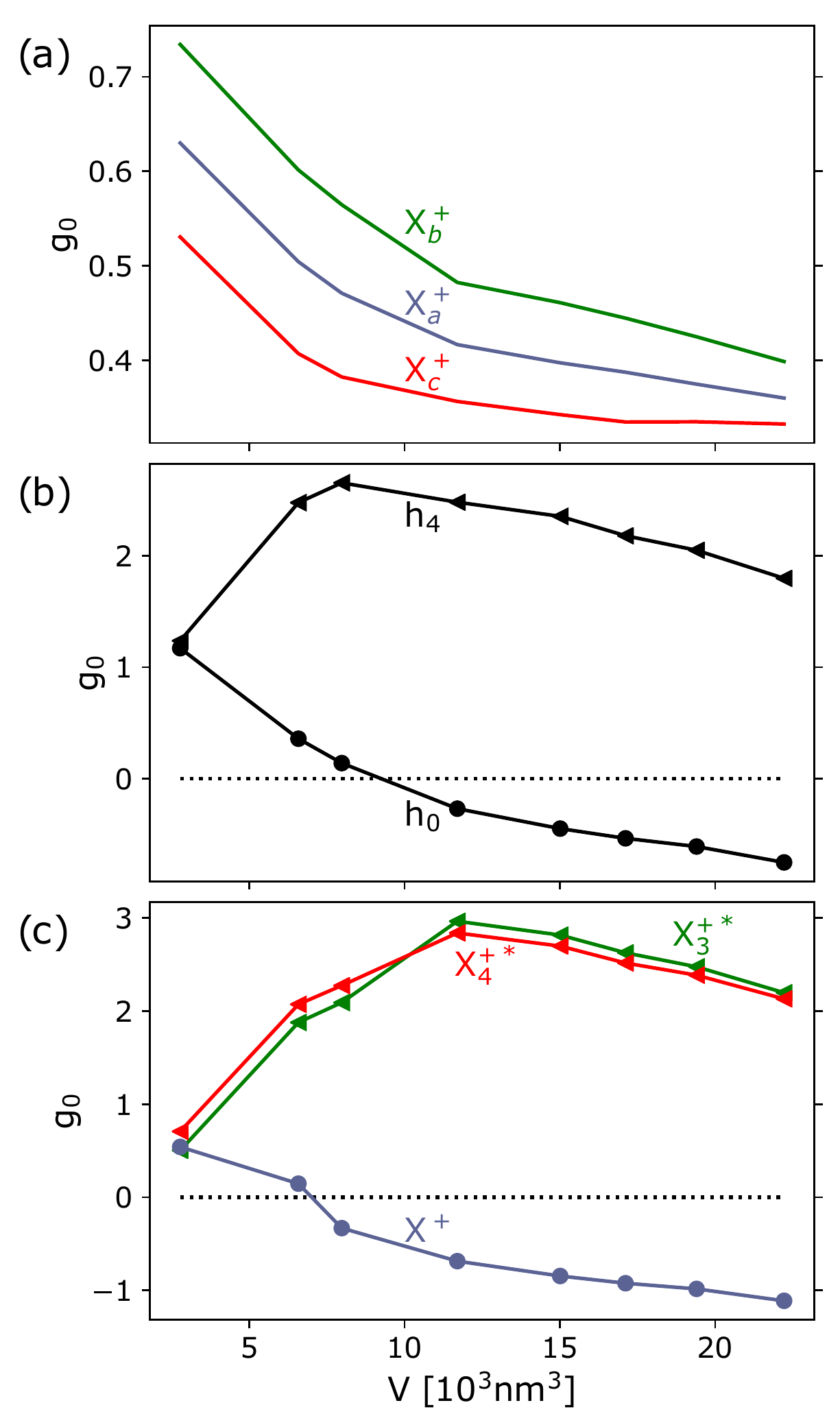}
	\caption{The QD volume dependence of the $g$-factor for (a) non-recombined positive trion states, (b) final SP hole states, and (c) emitted final single photon states. The colours identify the non-recombined trion states and the markers the final SP hole states.} 
	\label{figB.3}
\end{figure}

We now discuss the $g$-factors of X$^-$, X$^+$, X$_3^{+*}$, and X$_4^{+*}$, see Fig.~\ref{figB.1}~(b). Since $g_0$ of trions is given by the SP state the energy of which is subtracted during the recombination, we observe significantly faster decrease for X$^+$ ($b = 19$) than for X$^-$ ($b = 9$). The size dependence of subtracted SP electron (pink curve) and hole (orange curve) states are shown in the second column in Fig.~\ref{fig2}~(c). In the contrast with X$^0$, the $g$-factor of X$^+$ monotonically decreases towards negative values. The reason of smaller $g_0$ of X$^+$ is the larger content of $\ket{\rm LH}$ Bloch states. At the same time, the $g$-factors of the excited trions increase with increasing volume and for $V = 11.7\cdot 10^3\,\rm nm^3$ we observe a sudden decrease of those. In order to understand the aforementioned trends, we show in Fig.~\ref{figB.3} the QD volume dependence of $g$-factors of non-recombined positive trion states, the final SP hole state, and the final emitted single photon state. The colours identify the non-recombined trion states and the markers the final SP hole states. As expected, we observe similar trends for final SP hole state and the emitted photons. On the other hand, the $g$-factors of non-recombined trion states have positive values and decrease in the whole considered range of QD volumes $V$.

\section{Conclusions}

We have studied the size dependencies of the diamagnetic coefficient and $g$-factor of X$^0$, X$^-$, and X$^+$ ground states, and X$_3^{+*}$ and X$_4^{+*}$ excited positive trion states, respectively, of GaAs/AlGaAs cone shaped quantum dots. The magnetic field was applied in the $[001]$ growth direction. The sizes of quantum dots were changed in three ways,~i.e, for fixed height (and variable base diameter), for fixed base diameter (and variable height), and for fixed aspect ratio (and variable volume). To find the origin of the observed trends, we decompose the calculations into four levels of precision: dependencies for (i) single-particle electron and hole states, (ii) non-interacting electron-hole pair, (iii) electron-hole pair constructed from the ground state of both quasiparticles and interacting via the Coulomb interaction (i.e. with minimal amount of correlation), and (iv) that including the effect of correlation. The calculated dependencies have reasonably well reproduced the experimental trends observed in Ref.~\cite{Lobl2019}. 

The diamagnetic coefficients of the correlated X$^0$ are found to be described sufficiently well by the single-particle approach for small dots. The increase of that with QD size is found to be mostly due to the single-particle electron states.
The case of excitonic $g$-factor is more complex. Here, we find a decrease of that with size which is due to holes. However, the multi-particle effects for both, diamagnetic coefficient and $g$-factor, are found to be non-negligible in the case of large, weakly confined quantum dots.
The exchange and direct Coulomb interactions increase the absolute value of the excitonic $g$-factor. At the same time, the effect of correlation decreases that. 
Furthermore, we find that also the slope of the dependencies is smaller when the exchange and direct Coulomb interactions are included.

We have also studied size dependencies of magneto-optical properties for charged trions. Here, the dependencies of the diamagnetic shifts and $g$-factors of ground states of those complexes were found to have similar evolution with QD size as that for ground state X$^0$. Only for the latter ($g$-factor) we found the dependencies to be shifted in magnitude which we identified to be the result of the subtraction of the final single-particle state, electron for X$^-$ or hole for X$^+$. Strikingly, the excited positively charged exciton states, X$_3^{+*}$ and X$_4^{+*}$, show noticeably different behavior. Namely, the anomalous (enormous) diamagnetic shift in the case of the former (latter). Moreover, in both of the aforementioned complexes the $g$-factor has non-monotonic dependence with size. We interpret the former (diamagnetic shift) observation to be due to singlet-triplet mixing in large weakly confining GaAs dots. On the other hand, the latter phenomenon ($g$-factor) is again explained as a result of the subtraction of the final single-particle hole state.

\section{Acknowledgements}

The authors are indebted to M. C. L\"{o}bl for allowing to reprint his experimental data from Fig.~2 of Ref.~\cite{Lobl2019} in order to compare with the theory presented here and A. Rastelli for fruitful discussions.
The authors would also like to thank P. Klapetek and his group for providing the computational resources, on which the ${\bf k}\cdot{\bf p}$ and CI algorithms were performed.
Furthermore, the authors are thankful to S. Covre Da Silva, A. Rastelli from Johannes Kepler University of Linz and P. Klapetek from the Czech Metrology Institute in Brno for allowing to reprint their AFM measurements in Fig.~\ref{fig1}.
The work was financed by the project CUSPIDOR has received funding from the QuantERA ERA-NET Cofund in Quantum Technologies implemented within the European Union's Horizon 2020 Programme. In addition, this project has received national funding from the Ministry of Education, Youth and Sports of the Czech Republic and funding from European Union's Horizon 2020 (2014-2020) research and innovation framework programme under grant agreement No 731473.
%
%
The work reported in this paper was (partially) funded by project EMPIR 17FUN06 Siqust. This project has received funding from the EMPIR programme co-financed by the Participating States and from the European Union’s Horizon 2020 research and innovation programme.


\bibliography{references}

\section*{Appendix I.}

Here we give the description of the CI method. Let us consider the excitonic complex $\ket{\rm M}$ consisting of $N_e$ electrons and $N_h$ holes. The CI method uses as a basis the Slater determinants (SDs) consisting of $n_e$ SP electron and $n_h$ SP hole states which are determined using the envelope function method based on $\mathbf{k}\cdot \mathbf{p}$ approximation. Obtained SP states read
\begin{equation}
    \Psi_{a_i}(\mathbf{r}) = \sum_{\nu\in\{s,x,y,z\}\otimes \{\uparrow,\downarrow\}} \chi_{a_i,\nu}(\mathbf{r})u^{\Gamma}_{\nu},
\end{equation}
where $u^{\Gamma}_{\nu}$ is the Bloch wave-function of s-like conduction band or p-like valence band at the center of the Brillouin zone, $\uparrow$/$\downarrow$ mark the spin and $\chi_{a_i,\nu}$ is the envelope function, where $a_i \in \{ e_i, h_i \}$.

The trial function of considered excitonic complex reads
\begin{equation}
    \ket{\rm M} =  \sum_{\mathit m=1}^{n_{\rm SD}} \mathit \eta_m \ket{D_m^{\rm M}}, \label{eq9}
\end{equation}
where $n_{\rm SD}$ is the number of SDs $\ket{D_m^{\rm M}}$ and $\eta_m$ is the constant which is looked for using the variational method. The $m$-th SD is~\cite{Klenovsky2017}
\begin{equation}
\ket{D_m^{\rm M}} = \frac{1}{\sqrt{N!}} \sum_{\tau \in S_N} \rm sgn \mathit(\tau) \phi_{\tau\{i_1\}}(\mathbf{r}_1) \phi_{\tau\{i_2\}}(\mathbf{r}_2) \dots \phi_{\tau\{i_N\}}(\mathbf{r}_N).
\end{equation}
Here, we sum over all permutations of $N := N_e + N_h$ elements over the symmetric group $S_N$. For the sake of notation convenience, we joined the electron and hole wave functions from which the SD is composed of, in the unique set $\{\phi_1, \dots, \phi_N\}_m := \{ \Psi_{e_j}, \dots,\Psi_{e_{j+N_e-1}}; \Psi_{h_k}, \dots,\Psi_{h_{k+N_h-1}} \}$, where $j \in \{1,\dots, n_e \}$ and $k \in \{1,\dots, n_h \}$. In similar fashion we join the positional vectors of electrons and holes $\{r_1, \dots, r_N\}:= \{ \mathbf{r}_{e_1}, \dots,\mathbf{r}_{e_{N_e}}; \mathbf{r}_{h_1}, \dots,\mathbf{r}_{h_{N_h}} \}$

Further, we solve the Schr\"{o}edinger equation 
\begin{equation}
H^{\rm{M}} \ket{\rm{M}} = E^{\rm{M}} \ket{\rm{M}} ,
\end{equation}
where $E^{\rm{M}}$ is the eigenenergy of excitonic state $\ket{\rm{M}}$ and $H^{\rm{M}}$ is the CI Hamiltonian which reads $H^{\rm{M}} = H_0^{\rm{M}} + \hat{V}^{\rm{M}}$, where $H_0^M$ represents the SP Hamiltonian and $\hat{V}^{\rm{M}}$ is the Coulomb interaction between SP states. The matrix element of $\hat{V}^{\rm{M}}$ reads~\cite{Klenovsky2019}
\begin{equation}
\begin{split}
    &\bra{D_n^{\rm M}}\hat{V}^{\rm{M}}\ket{D_m^{\rm M}} = \frac{1}{4\pi\epsilon_0} \sum_{ijkl} \iint d\mathbf{r} d\mathbf{r}^{\prime} \frac{q_iq_j}{\epsilon(\mathbf{r},\mathbf{r}^{\prime})|\mathbf{r}-\mathbf{r}^{\prime}|} \\
    &\times \{ \Psi^*_i(\mathbf{r})\Psi^*_j(\mathbf{r}^{\prime})\Psi_k(\mathbf{r})\Psi_l(\mathbf{r}^{\prime}) - \Psi^*_i(\mathbf{r})\Psi^*_j(\mathbf{r}^{\prime})\Psi_l(\mathbf{r})\Psi_k(\mathbf{r}^{\prime})\}.
\end{split}
\label{eq:CoulombMatrElem}
\end{equation}
Here $q_i$ and $q_j$ label the elementary charge $|e|$ of either electron ($-e$), or hole ($e$), and $\epsilon(\mathbf{r},\mathbf{r}^{\prime})$ is the spatially dependent dielectric function. Note, that the Coulomb interaction is treated as a perturbation. The evaluation of sixfold integral in Eq.~\eqref{eq:CoulombMatrElem} is performed using the Green's function method~\cite{Schliwa:09,Stier2000,Klenovsky2017,Klenovsky2019}
\begin{equation}
\begin{split}
    \nabla \left[ \epsilon(\mathbf{r}) \nabla \hat{U}_{ajl}(\mathbf{r}) \right] &= \frac{4\pi e^2}{\epsilon_0}\psi^*_{aj}(\mathbf{r})\psi_{al}(\mathbf{r}),\\
    V_{ij,kl} &= \bra{\psi_{bi}}\hat{U}_{ajl}\ket{\psi_{bk}},
\end{split}
\end{equation}
where $a,b \in \{e,h\}$ and $\nabla := \left( \frac{\partial}{\partial x}, \frac{\partial}{\partial y}, \frac{\partial}{\partial z} \right)^T$.

Finally, we note that relating to the ongoing discussion~\cite{Bester2008} about the nature of the dielectric screening in Eq.~\eqref{eq:CoulombMatrElem},~i.e., whether or not to set $\epsilon(\mathbf{r},\mathbf{r}^{\prime})$ to unity for the calculation of the exchange integral, we refer the reader to our previous work~\cite{Klenovsky2019}. There, the CI calculation of FSS of the exciton and the trion binding energies relative to the exciton were performed for InAs/GaAs lens shaped QDs. It was found, that while setting $\epsilon(\mathbf{r},\mathbf{r}^{\prime}) = 1$ for the exchange interaction led in the case of the CI basis using 2 SP electron and 2 SP hole states to realistic values of both FSS and binding energy, for larger basis the latter (binding energy of trion) increased to unreasonably large values. However, setting $\epsilon(\mathbf{r},\mathbf{r}^{\prime})$ to bulk values recovered the experimentally realistic values of both FSS and binding energy, regardless of the CI basis size.

\begin{figure*}[htbp]
	\centering
		\includegraphics[width=140mm]{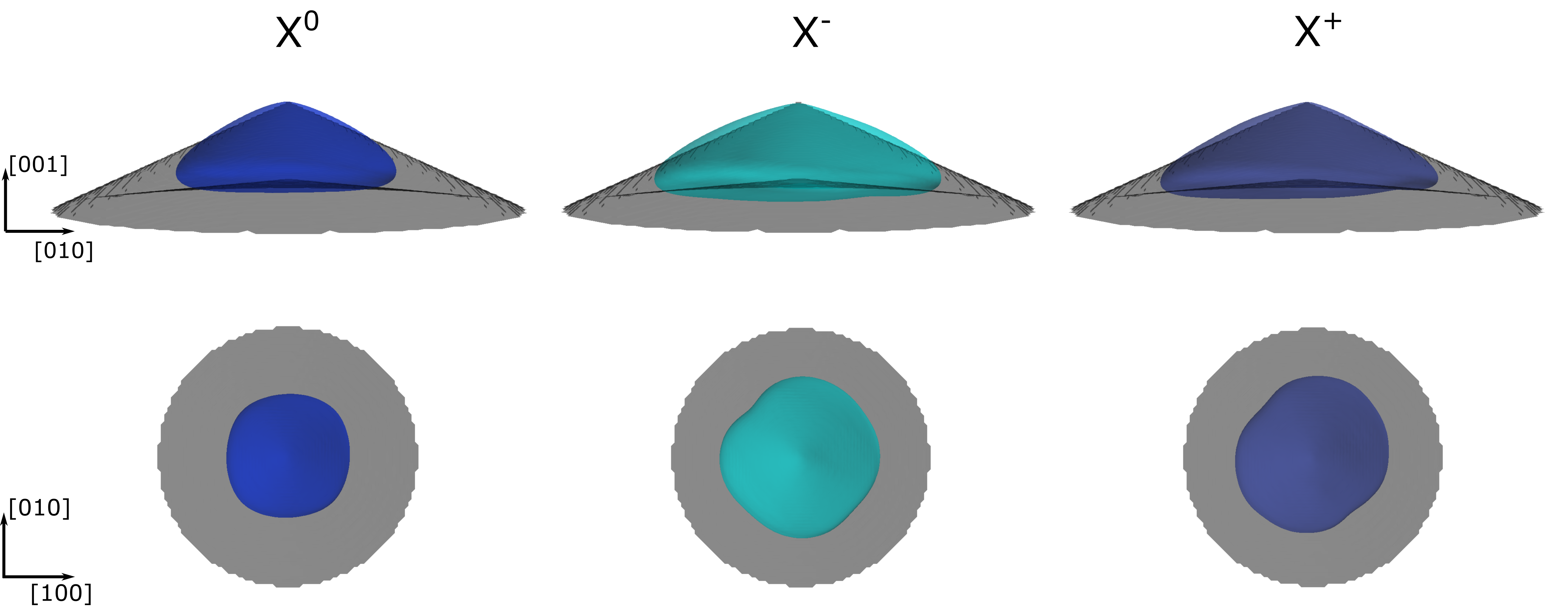}
	\caption{The isosurfaces of $90\,\%$ of total probability density of X$^0$, X$^-$, and X$^+$ ground states for $B_z = 0\,\rm T$, respectively. The upper row in (b) shows the side, while the bottom row the top view of the density, respectively. The insets in (b) indicate the crystallographic orientation of the cuts. Notice the larger volume of the probability densities for X$^-$ and X$^+$ in comparison with that for X$^0$.} 
	\label{figProbab}
\end{figure*}

\section*{Appendix II.}

To visualize the contents of SP states computed in multi-particle complexes calculated by CI, we need to transform the results of CI calculations to the basis of SP states instead of that of SDs. In this appendix we describe that method.

During the set-up of SDs, we create the matrix $\hat{A}$ with rank $n_{\rm SD} \times N$, where $m$-th row consists of SP states used in the corresponding SD
\begin{equation}
    A_m = \left( \Psi_{e_j}, \dots,\Psi_{e_{j+N_e-1}}; \Psi_{h_k}, \dots,\Psi_{h_{k+N_h-1}} \right).
\end{equation}

As a result of CI calculation we get $n_{\rm SD}$ eigenvectors (CI states) with $n_{\rm SD}$ components
\begin{equation}
    \ket{\rm M^{\mathit l}} = \left( \eta_1^l,\dots, \eta_{n_{\rm SD}}^{\mathit l} \right)^T,
\end{equation}
where index $l$ identifies the eigenvector. Now let us consider a particular SP state $\Psi_{e_j}$ $\{\Psi_{h_k}\}$. We choose those values of $\eta_m^l$ which corresponds to the $A_m$ consisting of $\Psi_{e_j}$ $\{\Psi_{h_k}\}$, sum the squares of the absolute values  
\begin{align}
    c_{e_j} &= \sum_m \sum_{j^{\prime}} |\eta^l_{m\,(j^{\prime})}|^2 \delta_{jj^{\prime}}, \\
    c_{h_k} &= \sum_m \sum_{k^{\prime}} |\eta^l_{m\,(k^{\prime})}|^2 \delta_{kk^{\prime}},
\end{align}
and obtain the vector
\begin{equation}
    \left( c_{e_1}^l,\dots, c_{e_{n_e}}^{l}; c_{h_1}^l,\dots, c_{h_{n_h}}^{l} \right)^T.
    \label{eqC.5}
\end{equation}
The values $c_{e_j}$ and $c_{h_k}$ are then normalized by imposing that $\sum_{j} c_{e_j}^l+\sum_{k} c_{h_k}^l = 1$. Since $|\eta_m^l|^2$ describes the weight of the corresponding SD in CI eigenvector, we look for the weights of individual SP electron or hole states. The example of the result is shown for X$^0$ in InAs/GaAs and GaAs/AlGaAs QD in Fig.~\ref{fig9} or for X$^+$ in GaAs/AlGaAs QD in Fig.~\ref{fig8}. 

The procedure described thus far allows us to study also other excitonic properties, such as the influence of multi-particle effects on band mixing or visualising the probability density of the studied excitonic complexes.

In the case of band mixing we multiply the contents of $\{\ket{\rm S}, \ket{\rm HH}, \ket{\rm LH}, \ket{\rm SO}\}$ of the particular SP state by the corresponding coefficient from Eq.~\eqref{eqC.5}. Hence, we get the matrix with rank $(n_e+n_h) \times 4$ for each $l$ and we sum separately all $\ket{\rm S}$, $\ket{\rm HH}$, $\ket{\rm LH}$ and $\ket{\rm SO}$ contents in that matrix to get the four corresponding values for each CI state. Finally, we normalize the contents in the same fashion as for Eq.~\eqref{eqC.5}.

On the other hand, for visualizing the probability density of an eigenstate of the complex $\ket{\rm M^{\mathit l}}$ with wave-function $\Phi_{\rm{M}}^l(\mathbf{r})$, we calculate
\begin{equation}
    |\Phi_{\rm{M}}^l(\mathbf{r})|^2 = \sum_{\mathit j} |c^l_{e_j} \Psi_{e_j} (\mathbf{r})|^2 + \sum_{\mathit k} |c^l_{h_k} \Psi_{h_k} (\mathbf{r})|^2.
\end{equation}
Similarly as before, the probability density is finally normalized,~i.e., $\braket{\rm M^{\mathit l}|M^{\mathit l}} = 1$. The example of the calculated probability density of X$^0$, X$^+$, and X$^-$ is shown in Fig.~\ref{figProbab}.

\section*{Appendix III.}
\label{appendIII}

Now we briefly describe the construction of the excited positive trion states. First, we introduce the single-particle approach considering the complex consisting of the electron in the ground state ($\uparrow_s$/$\downarrow_s$), the heavy hole in the ground state ($\Uparrow_s$/$\Downarrow_s$) and the heavy hole in the first excited state ($\Uparrow_p$/$\Downarrow_p$), where the total angular momentum of electron is $J = \pm 1/2$ and that of hole is $J= \pm 3/2$. Due to the exchange interaction, holes in the complex split into singlet and triplet states and, thus, the excited positive trions read~\cite{Igarashi2010,Huber2019}
\begin{equation}
        \ket{\rm X_{4}^{+*}}= \begin{cases} \uparrow_{s}(\Uparrow_{s}\Downarrow_{p}-\Downarrow_{s}\Uparrow_{p}) & J_z=+1/2 \\ \downarrow_{s}(\Uparrow_{s}\Downarrow_{p}-\Downarrow_{s}\Uparrow_{p}) & J_z=-1/2 \end{cases},
\label{eq2.10}
\end{equation} 
\begin{equation}
      \ket{\rm X_{3}^{+*}}= \begin{cases} \uparrow_{s}(\Uparrow_{s}\Downarrow_{p}+\Downarrow_{s}\Uparrow_{p}) & J_z=+1/2 \\  \downarrow_{s}(\Uparrow_{s}\Downarrow_{p}+\Downarrow_{s}\Uparrow_{p}) & J_z=-1/2  \end{cases},
      \label{eq2.11}
\end{equation} 
\begin{equation}
      \ket{\rm X_{2}^{+*}}=  \begin{cases}  \downarrow_{s}\Uparrow_{s}\Uparrow_{p} & J_z= +5/2 \\ \uparrow_{s}\Downarrow_{s}\Downarrow_{p} & J_z= -5/2 \end{cases},
\label{eq2.12}
\end{equation} 
\begin{equation}
\ket{\rm X_{1}^{+*}}= \begin{cases} \uparrow_{s}\Uparrow_{s}\Uparrow_{p} & J_z= +7/2\\ \downarrow_{s}\Downarrow_{s}\Downarrow_{p} & J_z= -7/2 \end{cases},
\label{eq2.13}
\end{equation}
where $J_z$ marks the projection of total angular momentum of the excited trions into the direction of $B_z$. We note that resulting from the introduced approximations, we are allowed to use the total angular momentum $J$ as a good quantum number here, instead of Kramers doublet. The singlet ($\rm X_{4}^{+*}$) and the triplet ($\rm X_{2}^{+*}$ and $\rm X_{3}^{+*}$) states emit single photon when s-shell electron and s-shell hole recombines. Hence, we get the energy of the emitted single photon when we subtract the energy of SP p-shell hole corresponding to the three SP electron and SP hole states where the s-shell electron--hole pair recombines. Note, that $\rm X_{1}^{+*}$ is the dark state due to the dipole selection rules. We stress, that since Eqs.~\eqref{eq2.10}--\eqref{eq2.13} do not include neither of multi-particle effects, we do not use them for the calculation discussed in the main text. However, single-particle approach gives us simple and clear picture of the approximate structure of the excited trion states. The calculation of energies considered in our study is discussed in the next paragraph.

\begin{figure*}[htbp]
	\centering
		\includegraphics[width=140mm]{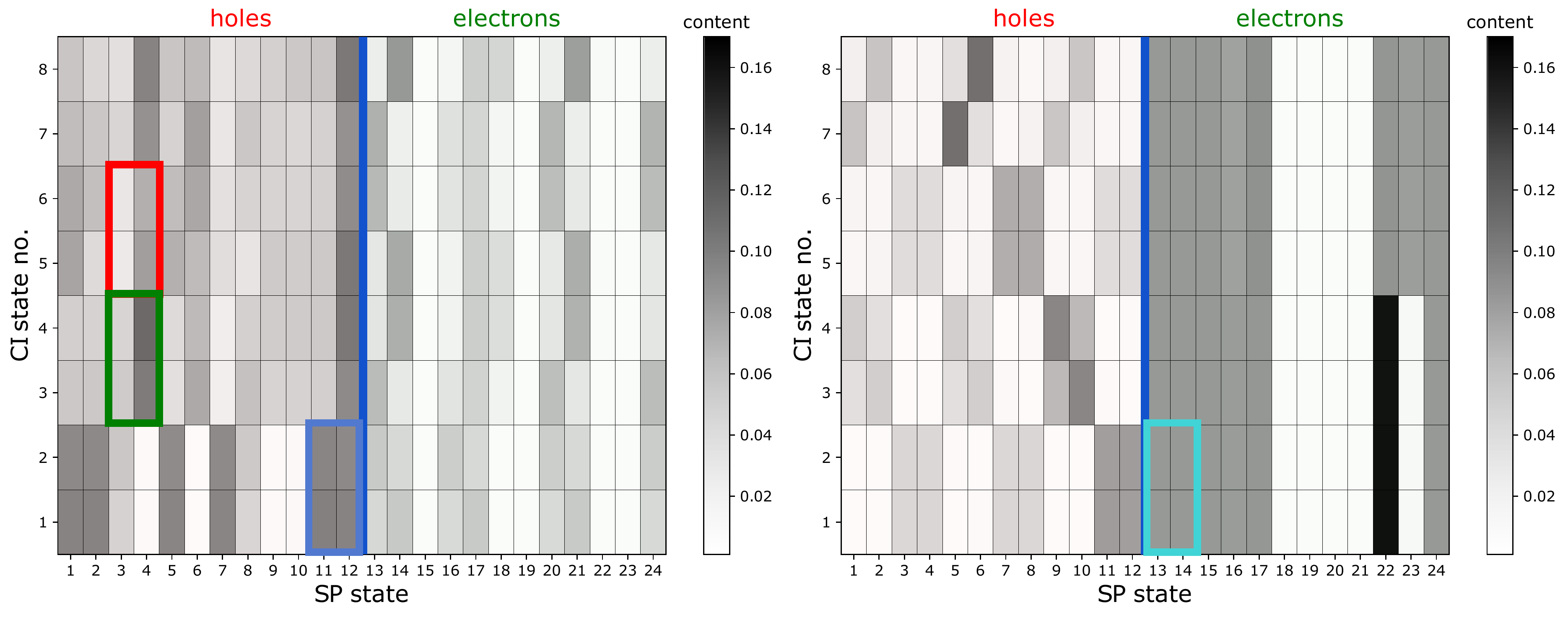}
	\caption{The contribution of SP states used for the basis of the CI calculations of QD with $d = 68\,\rm nm$ and $h = 16\,\rm nm$, in the non-recombined X$^+$ (left) and X$^-$ (right) trion states for $B = 0\,\rm T$. The CI calculations were performed with the basis of 12 SP electron (green) and 12 SP hole (red) states. The states in blue, green, and red rectangles in the left panel were used as the final SP states in constructing the studied emission energies of X$^+$, X$_3^{+*}$, and X$_4^{+*}$, respectively, and that in light blue of the right panel corresponds to the final SP state of X$^-$. The colors correspond to those in Fig.~\ref{figB.1}. Note that the CI states are ordered from 1 to 8 by their energy from lowest to highest. Note, that for the sake of clarity, in contrast with Fig.~\ref{fig9} we use linear scale here.} 
	\label{fig8}
\end{figure*}

Since each excitonic state calculated by CI consists of different amount of $n_e$ SP electron and $n_h$ SP hole states, we subtract the energy of the excited SP hole state which has the largest contribution in the considered CI state. That procedure allows us to reproduce the measured results~\cite{Huber2019}. In Fig.~\ref{fig8} we show SP hole states, the energies of which were subtracted in our investigation of X$^+$, $\rm X_{3}^{+*}$, and $\rm X_{4}^{+*}$ transitions. As can be seen, the largest contribution in the first and the second excited trion eigenstates (corresponding to numbers $3,4$ and $5,6$, and we mark them as $\rm X_b^+$ and $\rm X_c^+$, respectively) relates to the fourth excited SP hole state (that corresponds to the numbers $3,4$ and we mark that as $h_4$). Hence, the energies of the emitted single photons $\mathcal{E}$ which are the results of such transitions are
\begin{equation}
        \mathcal{E}_4\left({\rm X}_{4}^{+*}\right)= \begin{cases} E^{{\rm X}^+}_{c}\left\{u\right\} - E_4\left\{\Uparrow\right\}\\ 
        E^{{\rm X}^+}_{c}\left\{d\right\} - E_4\left\{\Downarrow\right\}  \end{cases},
\label{eq2.18}
\end{equation}
\begin{equation}
        \mathcal{E}_4\left({\rm X}_{3}^{+*}\right)= \begin{cases} E^{{\rm X}^+}_{b}\left\{u\right\} - E_4\left\{\Uparrow\right\}\\ 
        E^{{\rm X}^+}_{b}\left\{d\right\} - E_4\left\{\Downarrow\right\}  \end{cases},
\label{eq2.19}
\end{equation}
\begin{equation}
        \mathcal{E}_0\left({\rm X}^{+}\right)= \begin{cases} E^{{\rm X}^+}_{a}\left\{u\right\} - E_0\left\{\Uparrow\right\}\\ 
        E^{{\rm X}^+}_{a}\left\{d\right\} - E_0\left\{\Downarrow\right\}  \end{cases},
\label{eq2.20}
\end{equation}
where subscripts $a$, $b$, $c$ mark the Kramers doublet of the trion state before recombination ($a$ identifies the lowest energy). Furthermore, subscripts $0$ and $4$ label the Kramers doublet of the final SP hole state ($0$ identifies the energy of the ground state), and $u\,\,\{d\}$ denotes the higher $\{$lower$\}$ energy of considered trion doublet. For further details see also Ref.~\cite{Huber2019}.

\end{document}